\documentstyle[psfig,onecolumn]{mn}
\newif\ifAMStwofonts

\newcommand{\etal}{{\it et al.} }

\newcommand{\feka}{{Fe~K$\alpha$} }
\newcommand{\fekap}{{Fe~K$\alpha$}}

\newcommand{\fekalfap}{{Fe~K$\alpha$}}
\newcommand{\fekb}{{Fe~K$\beta$} }
\newcommand{\fekbp}{{Fe~K$\beta$}}
\newcommand{\fekbeta}{{Fe~K$\beta$} }

\newcommand{\nika}{{Ni~K$\alpha$} }
\newcommand{\nikap}{{Ni~K$\alpha$}}

\newcommand{\nh}{$N_{\rm H}$ }
\newcommand{\nhp}{$N_{\rm H}$}

\newcommand{\thetaobs}{{$\theta_{\rm obs}$} }
\newcommand{\thetaobsp}{{$\theta_{\rm obs}$}}

\newcommand{\dlc}{{$\Delta\lambda/\lambda_{\rm C}$} }
\newcommand{\dlcp}{{$\Delta\lambda/\lambda_{\rm C}$}}

\newcommand{\ein}{{$E_{\rm in}$} }
\newcommand{\einp}{{$E_{\rm in}$}}

\newcommand{\tablecosrange}{{Table~1} }
\newcommand{\tablecosrangep}{{Table~1}}

\newcommand{\figtorusgeomp}{{Fig.~1}}

\newcommand{\fignescvsalbp}{{Fig.~2}}
\newcommand{\figerange}{{Fig.~3} }

\newcommand{\figcmpalbedop}{{Fig.~4}}
\newcommand{\figexspecfigs}{{Fig.~5} }
\newcommand{\figexspecfigsp}{{Fig.~5}}
\newcommand{\figscatvstheta}{{Fig.~6} }

\newcommand{\figcmppexrav}{{Fig.~7} }

\newcommand{\figewvsnh}{{Fig.~8} }
\newcommand{\figewvsnhp}{{Fig.~8}}
\newcommand{\figewratvsnh}{{Fig.~10} }

\newcommand{\figewvstheta}{{Fig.~9} }
\newcommand{\figewvsthetap}{{Fig.~9}}
\newcommand{\figcmpcsa}{{Fig.~11} }

\ifoldfss
  \ifCUPmtlplainloaded \else
    \NewTextAlphabet{textbfit} {cmbxti10} {}
    \NewTextAlphabet{textbfss} {cmssbx10} {}
    \NewMathAlphabet{mathbfit} {cmbxti10} {} 
    \NewMathAlphabet{mathbfss} {cmssbx10} {} 
  \fi
  \ifAMStwofonts
    \ifCUPmtlplainloaded \else
      \NewSymbolFont{upmath} {eurm10}
      \NewSymbolFont{AMSa} {msam10}
      \NewMathSymbol{\upi}     {0}{upmath}{19}
      \NewMathSymbol{\umu}     {0}{upmath}{16}
      \NewMathSymbol{\upartial}{0}{upmath}{40}
      \NewMathSymbol{\leqslant}{3}{AMSa}{36}
      \NewMathSymbol{\geqslant}{3}{AMSa}{3E}

       \let\ge=\geqslant
    \fi
  \fi
\fi 

\ifnfssone
  \newmathalphabet{\mathit}
  \addtoversion{normal}{\mathit}{cmr}{m}{it}
  \addtoversion{bold}{\mathit}{cmr}{bx}{it}
  \newmathalphabet{\mathbfit} 
  \addtoversion{normal}{\mathbfit}{cmr}{bx}{it}
  \addtoversion{bold}{\mathbfit}{cmr}{bx}{it}
  \newmathalphabet{\mathbfss} 
  \addtoversion{normal}{\mathbfss}{cmss}{bx}{n}
  \addtoversion{bold}{\mathbfss}{cmss}{bx}{n}
  \ifAMStwofonts
    \ifCUPmtlplainloaded \else
      \UseAMStwoboldmath
      \makeatletter
      \new@mathgroup\upmath@group
      \define@mathgroup\mv@normal\upmath@group{eur}{m}{n}
      \define@mathgroup\mv@bold\upmath@group{eur}{b}{n}
      \edef\UPM{\hexnumber\upmath@group}
      \new@mathgroup\amsa@group
      \define@mathgroup\mv@normal\amsa@group{msa}{m}{n}
      \define@mathgroup\mv@bold\amsa@group{msa}{m}{n}
      \edef\AMSa{\hexnumber\amsa@group}
      \makeatother
      \mathchardef\upi="0\UPM19
      \mathchardef\umu="0\UPM16
      \mathchardef\upartial="0\UPM40
      \mathchardef\leqslant="3\AMSa36
      \mathchardef\geqslant="3\AMSa3E

       \let\ge=\geqslant
    \fi
  \fi
\fi 

\ifnfsstwo
  \DeclareMathAlphabet{\mathbfit}{OT1}{cmr}{bx}{it}
  \SetMathAlphabet\mathbfit{bold}{OT1}{cmr}{bx}{it}
  \DeclareMathAlphabet{\mathbfss}{OT1}{cmss}{bx}{n}
  \SetMathAlphabet\mathbfss{bold}{OT1}{cmss}{bx}{n}
  \ifAMStwofonts
    \ifCUPmtlplainloaded \else
      \DeclareSymbolFont{UPM}{U}{eur}{m}{n}
      \SetSymbolFont{UPM}{bold}{U}{eur}{b}{n}
      \DeclareSymbolFont{AMSa}{U}{msa}{m}{n}
      \DeclareMathSymbol{\upi}{0}{UPM}{"19}
      \DeclareMathSymbol{\umu}{0}{UPM}{"16}
      \DeclareMathSymbol{\upartial}{0}{UPM}{"40}
      \DeclareMathSymbol{\leqslant}{3}{AMSa}{"36}
      \DeclareMathSymbol{\geqslant}{3}{AMSa}{"3E}

       \let\ge=\geqslant
    \fi
  \fi
\fi 

\ifCUPmtlplainloaded \else
  \ifAMStwofonts \else 
    \def\upi{\pi}
    \def\umu{\mu}
    \def\upartial{\partial}
  \fi
\fi

\title{An X-ray Spectral Model for Compton-Thick Toroidal Reprocessors}
\author[K. D. Murphy \& T. Yaqoob]
       {Kendrah D. Murphy$^{1,2}$ and Tahir Yaqoob$^{2,3}$ \\
        $^1$MIT Kavli Institute for Astrophysics and Space Research, 77 Massachusetts
	Avenue, NE80-6013, Cambridge, MA 02139. \\
	$^2$Department of Physics and Astronomy, Johns Hopkins University, Baltimore, MD 21218. \\
	$^3$Astrophysics Science Division, NASA/Goddard Space Flight Center, Greenbelt, MD 20771.}
\date{
      Received }

\pagerange{\pageref{firstpage}--\pageref{lastpage}}
\pubyear{2009}

\begin{document}

\maketitle

\label{firstpage}

\begin{abstract} 

The central engines of both type 1 and type 2 AGNs are thought to harbor a toroidal structure
that absorbs and reprocesses high-energy photons from the central X-ray source. Unique features
in the reprocessed spectra can provide powerful physical constraints on the geometry, column
density, element abundances, and orientation of the circumnuclear matter. If the reprocessor is
Compton-thick, the calculation of emission-line and continuum spectra that are suitable for
direct fitting to X-ray data is challenging because the reprocessed emission depends on the
spectral shape of the incident continuum, which may not be directly observable. We present new
Monte-Carlo calculations of Green's functions for a toroidal reprocessor that provide significant
improvements over currently available models. The Green's function approach enables the
construction of X-ray spectral fitting models that allow arbitrary incident spectra as part of
the fitting process. The calculations are fully relativistic and have been performed for column
densities that cover the Compton-thin to Compton-thick regime, for incident photon energies up to
500~keV. The Green's function library can easily be extended cumulatively to provide models that
are valid for higher input energies and a wider range of element abundances and opening angles of
the torus. The reprocessed continuum and fluorescent line emission due to Fe~K$\alpha$,
Fe~K$\beta$, and Ni~K$\alpha$ are treated self-consistently, eliminating the need for {\it
ad~hoc} modeling that is currently common practice.  We find that the spectral shape of the
Compton-thick reflection spectrum in both the soft and hard X-ray bands in our toroidal geometry
is different compared to that obtained from disk models.  A key result of our study is that a
Compton-thick toroidal structure that subtends the same solid angle at the X-ray source as a disk
can produce a reflection spectrum that is $\sim 6$ times weaker than that from a disk.  This
highlights the widespread and erroneous interpretation of the so-called ``reflection-fraction''
as a solid angle, obtained from  fitting disk-reflection models to Compton-thick sources without
regard for proper consideration of geometry.  

\end{abstract}


\section{Introduction}
\label{torusintro}

The importance and wider implications of the absorption and reprocessing of high-energy radiation (X-rays to
$\gamma$-rays) in Compton-thick Active Galactic Nuclei (AGNs) have been recognized since the 1980s (e.g.
Makishima 1986; Guilbert \& Rees 1988; Lightman \& White 1988; Setti \& Woltjer 1989; Madau, Ghisellini \&
Fabian 1993, 1994; Matt \& Fabian 1994; Ikeda, Awaki, \& Terashima 2009). A class of type~2 active galactic
nuclei are known to be heavily obscured  by Compton-thick matter (e.g. Awaki \etal 1991; Done, Madejski, \&
Smith 1996;  Matt \etal 2000; Guainazzi, Matt, \& Perola 2005; Levenson \etal 2002, 2006).  The class of AGN
known as ULIRGs (ultra-luminous infrared galaxies) are also often characterized by heavy obscuration (e.g.
Braito \etal 2004; Teng \etal 2009 and references therein).  The obscuring matter in some
broad-absorption-line quasars (BALs) may also be Compton-thick (e.g. Gallagher \etal 2006; Chartas \etal
2007).  It is thought that even type~1 AGNs harbor a large, parsec-scale Compton-thick structure out of the
line-of-sight (e.g. Antonucci \& Miller 1985; Matt \etal 2000 and references therein). Thus, it is possible
that {\it most} AGN, regardless of their classification, harbor a toroidal structure that may or may not be
Compton-thick.  According to AGN unification schemes, (Compton-thick or Compton-thin) type 2 AGNs are objects
in which the line-of-sight to the central source passes through the reprocessing torus, and type 1 AGNs are
those for which the line-of-sight does not intercept this structure.  In addition to AGNs, heavy obscuration
has also been observed in X-ray binaries (XRBs), particularly in wind-fed X-ray pulsars (e.g. White, Nagase,
\& Parmar 1995; Rea \etal 2005, and references therein). 

Comparison of the observed spectrum from a source harboring a toroidal X-ray reprocessor with a
theoretically calculated spectrum can yield important constraints on the geometry, the solid angle
subtended at the radiation source, the element abundances, and the characteristic column density (or
Thomson depth) of the reprocessor.  The reprocessor modifies and distorts an incident spectrum
through photoelectric absorption and Compton scattering.  In particular, the reprocessor produces a
characteristic hump in the spectrum between $\sim 10-30$~keV and diminishes the intrinsic continuum
below $\sim 10$ keV.  Additionally, the structure may produce observable fluorescent emission lines,
as well as so-called ``Compton shoulders" on the emission lines from the most abundant elements (e.g.
Ghisellini, Haardt, \& Matt 1994; Done \etal 1996; Matt 2002; Watanabe \etal 2003).  


The reflection continuum in AGNs likely includes contributions from both the putative accretion disk as well as a
more distant X-ray reprocessing structure.  It is for this reason that the fluorescent line emission from AGNs is
often complex.  For example, a narrow Fe~K line (FWHM $< 5000 \rm \ km \ s^{-1}$ or so) at $\sim6.4$ keV is
ubiquitous in observations of AGNs (e.g. Yaqoob \& Padmanabhan 2004).  This narrow line component likely
originates from circumnuclear matter far from the black hole and/or from the outer regions of the accretion
disk.  The observed peak energy of the narrow \feka line at 6.4 keV in many AGNs provides overwhelming evidence
that the narrow core of the \feka line in AGNs originates in cold matter (e.g. Nandra \etal 1997a,b; Sulentic
\etal 1998; Weaver, Gelbord, \& Yaqoob 2001; Reeves 2003; Page \etal 2004; Yaqoob \& Padmanabhan 2004;
Jim\'{e}nez-Bail\'{o}n \etal 2005;  Zhou \& Wang 2005; Jiang, Wang, \& Wang 2006; Levenson \etal 2006). A broad
Fe~K line (i.e., FWHM of the order of 30,000 $\rm km \ s^{-1}$ or more) is sometimes detected as well (e.g.
Nandra \etal 2007).  The broad component most likely originates close to the black hole, in the accretion disk,
as it appears to be affected by gravitational as well as Doppler energy shifts.   

It is possible that some or all of the narrow Fe K emission line in AGN comes from the optical broad-line region
(e.g. see Bianchi et al. 2008). However, the size scale of the line emitter only affects the spectrum through
velocity broadening (in particular shaping the emission-line profiles), so our model is applicable to any
location in the central engine from the broad-line region and beyond since velocity broadening can be applied to
the Monte-Carlo results. Gaskell, Goosmann, \& Klimek (2008) argue that there is considerable observational evidence that the
broad-line region itself has a toroidal structure, and that there may be no distinct boundary between the
broad-line region and the classical parsec-scale torus.  Hereafter we shall refer to {\it any} toroidal
distribution of matter in the central engine as ``the torus'', regardless of it's actual size or physical
location in the central engine.

Duly taking into account and modeling possible additional reflection features from an accretion disk, one might be
able to obtain constraints on the column density of the putative torus.  Some type 2 AGNs that appear to be
Compton-thin may actually harbor a Compton-thick torus that has an inclination angle such that the line-of-sight
passes near the surface of the torus.  Even in cases where the reprocessor is not intercepted directly along the
line-of-sight (e.g. a subclass of type~1 AGNs), Compton-reflection and fluorescent emission-line signatures may
still be imprinted on the observed (direct-intrinsic plus reflected) X-ray spectrum.  Therefore, one must model
reflection from a possible distant-matter component in type~1 systems, especially if narrow Fe~K line emission is
detected, in order to obtain correct constraints on other physical parameters, such as those pertaining to the
accretion disk.  Fluorescent emission lines from matter illuminated by X-rays provide additional diagnostics that
can be critical for constraining models, especially for breaking degeneracies that are present in continuum-only
fitting.  Estimating the true thickness out of the line-of-sight of the reprocessor requires full modeling of the
entire broad-band continuum and fluorescent line emission.  Spectral-fitting models based on the results described
here will facilitate such modeling, which will be applicable to {\it all} AGNs since it yields direct constraints
on the geometry and column density of distant matter out of the line-of-sight.

Our calculations provide several improvements over currently available models and highlight some hitherto
under-appreciated geometry-dependent effects that have important observational consequences.  In
\S\ref{prevwork} we give an overview of previous work and currently available models.  In \S\ref{greens} we
discuss our Monte-Carlo Green's function calculations in detail along with the critical assumptions made.  We
present the results of integrating these Green's functions for a simple input continuum in \S\ref{bbspec} and
compare them with a standard disk-reflection model.  We summarize our results and conclusions in
\S\ref{torusconcl}.

\section{Previous work and current modeling practices} 
\label{prevwork} 

Current modeling practices generally do not fully utilize all of the physical relationships between the
fluorescent lines and reprocessed continuum by imposing the appropriate constraints that can only be realized
with a model that treats both line and continuum components self-consistently.  The available fitting codes
incorporated in X-ray spectral fitting packages such as {\tt XSPEC} (Arnaud 1996) or {\tt ISIS} (Houck \&
Denicola 2000) have all been deficient in one or more aspect of modeling the complex transmission and
reflection spectrum.  For example, a common method of fitting involves ignoring scattering by the material
altogether, and simply including a line-of-sight absorption component.  A slightly better (but still
restrictive) approach is to consider simple attenuation by material in the line-of-sight, still neglecting
photons that may be reflected into the line-of-sight, as in the {\tt XSPEC} model {\tt cabs}.  Scattered photons are
simply discarded in {\tt cabs}, and there is no energy dependence for the scattering.  Another model, {\tt
plcabs}, calculates transmission and scattering for a spherical distribution of matter surrounding an X-ray
source.  This may be a more likely configuration (and yield a more realistic spectral fit), but it is valid
only up to $\sim 10-18$ keV in the rest frame, depending on the column density of the material.  Furthermore,
it is only valid for a column density of up to $5\times10^{24} \rm cm^{-2}$.  

A more commonly used fitting method is to attempt to imitate scattering in the torus using a disk-reflection
model (in particular, {\tt pexrav} -- see Magdziarz \& Zdziarski 1995).  This geometry is not physically
relevant and does not account for transmission through the material; therefore the derived parameters are not
useful.  In fact, in the present work we show that the commonly derived ``covering factor'' or ``solid angle''
from this model is severely geometry-dependent. In addition, as with {\tt cabs} and {\tt plcabs}, the model does
not treat the Fe~K line emission.  The line-emission component must then be included {\it ad~hoc} and so does not
incorporate constraints set by other key parameters of the matter from which it originates.  Furthermore, the
model parameters from disk reflection cannot be related to the line-of-sight column density.  Clearly a more
sophisticated, less {\it ad~hoc} method of spectral fitting is required if we hope to truly understand the large-scale
structure of AGN and constrain unification models.

We note that, even for poorer quality data, Compton scattering must be included when modeling heavily obscured
sources in order to derive the correct column density and intrinsic luminosity, even if the detailed continuum
curvature and/or emission-line shape is not well-defined by the data. Erroneous absorption-only model-fitting
(with disk-reflection often used to mimic the scattered continuum) is abundant in the literature, mainly because a
correct model has not been available.  

Ideally, one should  perform iterative spectral fitting to an observed X-ray spectrum, minimizing a fit statistic
to find the best-fitting parameters of the reprocessor model for a given data set. However, a spectral-fitting
model for Compton-thick reprocessors that is valid for arbitrary incident spectra and for energies high enough to
sample the full range of reprocessing features has not previously been available.  The reprocessed spectrum must,
in general, be calculated by means of Monte-Carlo methods. This is prohibitively slow for iterative spectral
fitting because a model needs to be calculated hundreds to thousands of times per second, as the model parameters
are varied to find the best  fit. In such cases it is customary to create a pre-calculated grid of spectra (for
sets of parameter values of interest) and use interpolation for fast, {\it in-situ} calculation of a model,
updating input parameters at every iteration of the fit (and during calculation of statistical errors).  Clearly
there are limitations on the dimensionality and size of the model grid that may ultimately impose limitations on
the  versatility of the model and on the accuracy of interpolation. Recent Monte Carlo calculations presented by
Ikeda \etal (2009) provide a significant advance in terms of  self-consistently modeling the continuum and
Fe~$K\alpha$ emission line from Compton-thick matter. However, each Monte Carlo run corresponds to a specific
incident spectrum (not a Green's function), Compton scattering is treated using the non-relativistic Thomson
differential cross-section in the relativistic regime, and the models are calculated only up to 100~keV. In
addition, the spectra in the soft X-ray band (below 6~keV)  are adversely affected by small-number statistics.

For a Compton-scattering medium the reprocessed spectrum actually depends on the shape of the incident
spectrum, so that one is then forced to use a grid with a fixed spectral shape of the incident spectrum. 
This is unsatisfactory because the incident spectral shape is generally not known {\it a priori} and should
be obtained from the spectral fitting process itself. The simplest observationally-relevant incident spectra
will generally have at least two non-trivial parameters: for example, a power-law spectral index and
high-energy cut-off, or an optical depth and temperature for a Comptonizing plasma  (e.g. Haardt \& Maraschi
1991; Ghisellini \etal 1994; Zdziarski, Poutanen, \& Johnson 2000; Petrucci \etal 2001).  The Green's
function approach for solving for the radiative transfer problem for arbitrary incident spectra has been
well-documented (e.g. Illarionov \etal 1979; Lightman, Lamb, \& Rybicki 1981; White, Lightman, \& Zdziarski
1988; Magdziarz \& Zdziarski 1995). Whilst an implementation for a spectral-fitting tool in the case of a
semi-infinite medium is available  (a Green's function version of {\tt pexrav}), our results will  now
facilitate the construction of spectral-fitting models for a finite medium.  Additionally, our model will
improve upon other so-called ``table models" that exist to date by treating {\it both} the Fe~K line and the
continuum with sufficient detail to be applicable to the high-resolution data from current and planned X-ray
missions.  One particular advantage of our model is that it allows us to study inclination angle effects in
more detail than previously possible with other reprocessor models.  Furthermore, our model has superior
low-energy statistics relative to comparable reprocessor models.  All of these features therefore render our
model highly suitable for spectral fitting of real data. 

Our model geometry is also unique. Previous work has considered geometries that are spherical (e.g.
Leahy \& Creighton 1993; Matt, Pompilio, \& La Franca 1999), spherical-toroidal\footnote{A sphere with
a bi-cone removed at the poles.} (e.g. Ghisellini \etal 1994; Ikeda \etal 2009), and toroidal with a
rectangular cross-section (e.g. Awaki \etal 1991; Krolik, Madau, \& \.{Z}ycki 1994).  The present work,
as far as we know, is the first to investigate a three-dimensional torus with a circular cross-section
in this context. 

\section{Green's Functions}
\label{greens}

We have constructed a Monte-Carlo code to calculate grids of Green's functions to model the passage of X-rays
through a distant, toroidal reprocessor.  The Green's functions are created by firing mono-energetic photons
into the structure (see \S\ref{geometry}) and tracing each photon until it is absorbed or escapes the
reprocessor.  There is a given probability of interaction with the medium through Compton scattering or
absorption, which is determined by the respective cross-sections (see \S\ref{csscater}), the Compton depth
(or column density) in the direction of photon propagation, and element abundances.  If a photon is absorbed
by a particular atom and has an energy greater than the K~edge energy for that atom there is a further
probability that a new photon will be emitted by K-shell line fluorescence (\S\ref{felines}).  Each time the
photon is scattered, or when a new line photon is emitted, it is assigned a new direction and energy (see
\S\ref{gridintervals}).  The distance and the probability to escape the structure is then calculated, and the
process is repeated until the photon is absorbed (but does not produce a fluorescent emission line) or
escapes without interacting with the torus again.  If a photon escapes the structure and is traveling in a
direction such that it intercepts the structure again, it is allowed to re-enter and resume interaction with
the medium.  The photons that finally escape the structure completely are flagged as continuum or as line
photons of a particular atomic species.  

Each escaping photon has a final energy and direction of propagation.  The photons are sorted into a
two-dimensional grid of Green's functions of energy (or wavelength) and escape direction, one grid
for continuum photons and one for each of type of K-shell emission line.  For a given
escape-direction bin, the one-dimensional array of photons  versus energy (or wavelength) bins
constitutes a Green's function.

\subsection{Assumed Geometry \& Structure}
\label{geometry}

We consider a tube-like, azimuthally-symmetric torus (see \figtorusgeomp).  Here $c$ is the distance from
the center of the torus (located at the origin of coordinates) to the center of the ``tube" and $a$ is
the radius of the tube, but note that only the {\it ratio}, $c/a$, is important for our calculations. 
This corresponds to the classic ``doughnut" type of geometry for the obscuring torus which is a key
component of AGN unification schemes.  

\begin{figure}
\centerline{
	\psfig{figure=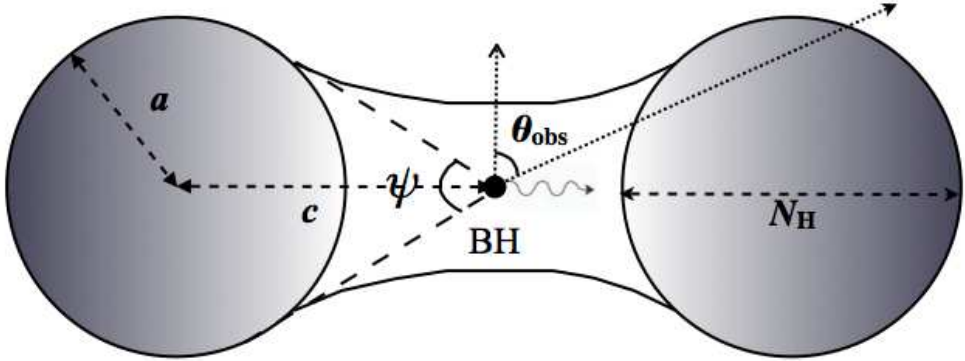,height=6cm}}
	\caption{\footnotesize Assumed model geometry. The half-opening angle is
	given by $\frac{\pi-\psi}{2}$ 
	and the inclination angle of the observer's line of sight with the symmetry axis of the
	 torus is given
	by \thetaobsp.  The equatorial column density, \nhp, of the torus is 
	defined by the diameter of the tube of the ``doughnut".}  
\end{figure}

The inclination angle between the observer's line of sight and the symmetry axis of the torus is given by
\thetaobsp, where \thetaobs$=0^{\circ}$ corresponds to a face-on observing angle and \thetaobs$=90^{\circ}$
corresponds to an edge-on observing angle.  

The equatorial column density, \nhp, is defined as the equivalent Hydrogen column density through the
diameter of the tube of the torus (as indicated in \figtorusgeomp).  The actual line-of-sight column density
is: \begin{equation} \label{eq:losnh}  N_{\rm H, \ l.o.s.} = N_{\rm H}\left[1-\left(\frac{c}{a}\right)^{2}
\cos^{2}{\theta_{\rm obs}}\right]^{\frac{1}{2}}.  \end{equation}   The mean column density, integrated over
all lines-of-sight through the torus, is $(\pi/4)$\nhp.  The column density may also be expressed in terms of
the Thomson depth: $\tau_{\rm T} = \frac{11}{9} N_{\rm H}\sigma_{\rm T} \sim 0.81N_{24}$ where $N_{24}$ is
the column density in units of $10^{24} \rm \ cm^{-2}$. Note that this assumes that the abundance of He is
10\% by number and that the number of electrons from all other elements aside from H and He is negligible. 
The mean number of electrons per H atom is $\frac{1}{2}(1+\mu)$, where $\mu$ is the mean molecular weight.
In this case, $\mu = 13/9$, for which  $\frac{1}{2}(1+\mu) = 11/9$.

The half-opening angle ($\frac{\pi-\psi}{2}$) of the torus is assumed to be $60^{\circ}$ (or $c/a=2$; see
\figtorusgeomp), corresponding to a scenario where there are equal numbers of unobscured and obscured AGN
in the universe.  The observed ratio of type~1 to type~2 AGN quoted in the literature varies widely (e.g.
Treister, Krolik, \& Dullemond 2008 reference values of type~1:type~2 AGN ratios that range from $\sim 2
: 3$ to $\sim 2 : 1$ and Elitzur 2008 refers to studies that find values of $\sim 3 : 7$ and $1 : 1$). 
Moreover, the ratio of type~1 to type~2 AGN is not necessarily the same as the ratio of unobscured to
obscured AGN.  The actual ratio is uncertain as it is dependent upon Seyfert classification definitions,
but we have chosen a value that is consistent with observational findings reported in the literature.  In
future work, our calculations will be extended to include a range of opening angles.  

We assume that the X-ray source emits isotropically and that the reprocessing material is uniform and
essentially neutral (cold).  Dynamics are not included in the Monte-Carlo code. Kinematic information can
be approximated by convolving the Green's functions or final output spectrum with a velocity function.   

Calculation of each Green's function involves multiple numerical calculations of the escape
distances after each scattering and/or line-emission event.  In order to calculate escape distances,
we derived a quartic equation, $Z$, whose zeros yield the desired torus-boundary values as a
function of the initial position and direction of a photon.  In our Monte-Carlo code, the zeros of
this equation were found using a bi-section method for a given tolerance, which we took to be
$10^{-2}$, such that the solution yields a value in the range of $Z=0\pm0.01$.  This tolerance value
was chosen to optimize accuracy and run-time of the Monte-Carlo code.  Note that the initial photon
position need not be inside the torus in order to obtain torus-boundary solutions, which is relevant
in our code for finding the initial entry points of photons that are first injected into the torus. 
The function, $Z=0$, yields up to four solutions and therefore the appropriate solution must be
selected in each case.  

The Green's functions have been computed over $N_{n}$ output bins of fractional Compton wavelength shift (in units
of $h/m_{e}c$, where $h$ is the Planck constant, $m_e$ is the rest mass of the electron, and $c$ is the speed of
light -- see \S\ref{gridintervals}), for $N_{m}$ incident energies, $E_{m}$, and $N_{i}$ values of the equatorial
column density through the structure, $N_{H,i}$.  The Monte-Carlo output was further sorted into $N_{k}$ angular
bins with cosine bin centers of $\mu_{k}$ (corresponding to the observer's orientation relative to the axis of the
torus).  Using equal cosine intervals for the inclination angle bins ensures equal solid angles for each bin, thus
ensuring that the numbers of photons in the bins with the smallest inclination angles are not photon-starved. 
Note that we do not need to retain the azimuthal escape angle information since the torus is axi-symmetric.  We
also utilize the polar symmetry about \thetaobs$=90^{\circ}$ by treating escape angles of \thetaobs and
$180^{\circ}-$\thetaobs as equivalent.  Thus there are $N_{k}$ Green's functions (a ``response'' versus $N_{n}$
output bins) for each $E_{m}$ and $N_{H,i}$.  Using interpolation, any of the Green's functions can then be
evaluated at any energy, and for any combination of \thetaobs and \nh on the grid.  Storing the Green's functions
as a function of Compton wavelength shift (instead of energy) is essential for interpolation because they have
sharp discontinuities at similar values of wavelength shift, not absolute energy (see \S\ref{gridintervals}). 

Therefore, for a given arbitrary input spectrum $F_{\rm in}(E_{l}, x_{1}, x_{2},...)$ , where 
$x_{1,2...}$ are the parameters of the input spectrum, the  transmitted and reprocessed
spectrum, $F_{\rm R}(E_{i})$, can be computed on an arbitrary energy grid by numerically
integrating the input spectrum over a  Green's function grid, interpolated on $E_{m}$, $G(E_{i},
E_{m}, N_{H,j}, \mu_{k}, g_{1}, g_{2},...)$ (where the $g_{1,2...}$ are the other parameters of
the Monte-Carlo model): \begin{eqnarray} \label{eq:greensfn} F_{\rm R}(E_{i}, x_{1}, x_{2},...,
g_{1}, g_{2},...)_{jk} & = & \sum_{(E_{l} \ge E_{i})}[G(E_{i}, E_{l}, N_{H,j}, \mu_{k}, g_{1},
g_{2},...) \ F_{\rm in}(E_{l}, x_{1}, x_{2},...) \Delta E_{l}].
\end{eqnarray}

\subsection{Zeroth-order Continuum}

The continuum Green's functions only include continuum photons that scattered at least once before escaping, and
do {\it not} include the zeroth-order (straight-through) continuum photons that were not absorbed and that
escaped without being scattered. The zeroth-order continuum for a given photon energy is simply \begin{eqnarray}
\label{eq:nzeroth}  N_{\rm 0^{th}} & = & N_{\rm in}e^{-(\tau_{\rm a}+\tau_{\rm s})}, \end{eqnarray} where
$\tau_{\rm a}$ and $\tau_{\rm s}$ are the absorption and scattering optical depths along the line-of-sight,
respectively, and $N_{\rm in}$ is the number of input photons per unit solid angle.  Therefore the zeroth-order
photon numbers can be computed numerically, independent of the Monte-Carlo results.  The unscattered
(zeroth-order) continuum can be added to the scattered continuum and the emission-line spectra calculated from
the Green's functions in order to obtain the net spectrum.

\subsection{Compton Scattering and Photoelectric Absorption}
\label{csscater}
A full exposition of Monte-Carlo methods for treating Compton scattering can be found in
Pozdnyakov, Sobol, \& Sunyaev (1983). The techniques for finite-Compton-depth transmission and
reflection models have much in common with Monte-Carlo X-ray pure-reflection codes (e.g. George
\& Fabian 1991; Reynolds \etal 1994). We used the full differential Klein-Nishina
Compton-scattering cross-section (used to generate the probability distribution to select the
angle between initial and final photon directions), and the full Klein-Nishina angle-averaged
Compton-scattering cross-section (which determines the probability of scattering for a given
initial photon energy). See, for example, Pozdnyakov \etal (1983) or Coppi and
Blandford (1990) for the full expressions.  The total Klein-Nishina cross-section drops rapidly
with increasing initial photon energy (relative to $m_{\rm e}c^{2}$) so the
corresponding optical depth to scattering drops.  Also, forward scattering dominates over
back-scattering more and more as the initial photon energy increases (in the Thomson regime
limit for low energies, forward and back-scattering are equally likely). In the electron rest
frame, Compton scattering changes the energy of the incident energy of a photon, $E_{i}$, to
$E_{f} = E_{i}/[1 + (E_{i}/m_{\rm e}c^{2})(1 - \cos{\alpha})]$, where $\alpha$ is the scattering
angle relative to the initial photon direction, drawn from the differential cross-section
distribution.  A new azimuthal angle (about the new photon direction) is drawn 
from a uniform probability distribution.  Since we are assuming that the torus is cold, the energy
shifts in the electron rest frame are taken to be the same as those in the lab frame.

We utilized photoelectric absorption cross-sections for 30 elements as described in Verner \& Yakovlev (1995) and
Verner \etal (1996).  Although these cross-section parameterizations are only valid up
to 100 keV, the absorption cross-sections near 100 keV and at higher energies can be approximated by a simple
power-law form and are orders of magnitude less than the values at the threshold energies (all $< 10$~keV). 
Therefore, we extrapolated the total cross-section for energies above 100~keV using a power law with a slope
equal to that in the $90-100$~keV interval.  We used Anders and Grevesse (1989) elemental cosmic abundances in
our calculations.

\subsection{Fluorescent Emission Lines} \label{felines} We included fluorescent line emission in our
Monte-Carlo code for Fe and Ni. Fluorescent lines from other cosmically-abundant elements (such as
C, O, Ne, Mg, Si, and S) are less observationally relevant than those from Fe and Ni due to
their small fluorescence yield and because lower-energy line photons have a greater probability of
being absorbed before escaping the medium than higher-energy photons. See for example, Reynolds
\etal (1994), and Matt, Fabian, \& Reynolds (1997), who have calculated disk-reflection spectra,
including all of these fluorescent lines. Although \nika line emission has been detected
in a few sources (e.g. Molendi, Bianchi, \& Matt 2003, and references therein),  the fluorescent
lines from Fe are by far the most important, due to the high fluorescence yield of Fe, and the
relatively high abundance of Fe (the Ni solar abundance is only $\sim 3-6$\% of the Fe abundance). 
Nevertheless, the fluorescent lines due to lighter elements will be included in future work since
they can be important in ``pure-reflection'' spectra. 

We therefore carried out a detailed treatment of the Fe K lines, explicitly including \fekbeta (at
7.058~keV) in addition to \fekalfap. For neutral Fe, the K$\alpha$ emission consists of two lines,
K$\alpha_{1}$ at 6.404 keV and K$\alpha_{2}$ at 6.391 keV, with a branching ratio of 2:1 (e.g. see
Bambynek \etal 1972).  We treated these as a single line, but the two lines can easily be
reconstructed from our Monte Carlo results.  We additionally accounted for \nika emission at 7.472
keV (see Bearden 1967).  We assumed Fe and Ni fluorescence yields of 0.347 and 0.414, respectively
(see Bambynek \etal 1972) and an \fekb to \feka ratio of 0.135 (representative of the range of
experimental and theoretical values discussed in Palmeri \etal 2003).  These values, as well as the
line energies, are subject to measurement and theoretical uncertainties.  Additional discussion and
a collection of results in the literature for these atomic parameters (including line
energies) can be found in Kallman \etal (2004).

Our Monte-Carlo code calculates, using the relevant cross-sections, the probabilities for the photon to be
absorbed by either an Fe~K or Ni~K shell, and in turn the probability for a fluorescent line to be created.  We
used K-shell absorption cross-sections and energies calculated by Verner \etal (1996) to calculate these
probabilities.  If Fe~K shell fluorescence occurs, it is then determined if the emitted photon is an \feka or \fekb
line photon.  Each line photon is ``labeled", identifying it explicitly as \fekap, \fekbp, or
\nikap.  Once a fluorescent line photon is emitted, its passage through the medium is followed as it would be
for any other photon (i.e., it is assigned a fresh starting energy and absorption, scattering, and escape are
treated until the photon is absorbed or escapes).  The Monte-Carlo output for the escaping fluorescent emission
line photons consists of a single number for the zeroth-order line photons and an array containing the energy
distribution of the scattered line photons (i.e., the Compton shoulder).  Note that since the total
photoelectric absorption cross-section at the energy of the \fekb line is less than that at the energy of the
\feka line, the observed intensity ratio of \fekb to \feka may be larger than the \fekbp/\feka branching
ratio.

\subsection{Grid Parameters}
\label{gridintervals}

In the present work, we calculated the Green's functions for 28 values of the column density parameter, \nhp, in
the range $10^{22} \rm \ cm^{-2}$ to $10^{25} \rm \ cm^{-2}$.  The range in column density corresponds to a range
in the Thomson depth of $\sim 0.0081 - 8.1$. The lower end of this range, where scattering becomes less
important, ensures that there is no significant discontinuity when switching between spectra generated from the
Green's functions and those calculated from standard absorption models that do not include Compton scattering.  

The output Green's functions are stored in 10 \thetaobs bins that are equal in solid angle (see \tablecosrangep). 
For  the current geometry with a half-opening angle of $60^{\circ}$, this corresponds to 5 bins that have
lines-of-sight that intercept the torus (with \thetaobs between $60^{\circ} \rm \ and \ 90^{\circ}$) and 5 that
have lines-of-sight that do not intercept the torus (with \thetaobs $<60^{\circ}$).

\begin{table}
\caption{\thetaobs Bin Boundaries.  Bins 1--5 correspond to lines-of-sight that
do not intercept the torus for the opening angle discussed here.  Bins 6--10
correspond to lines-of-sight that intercept the torus.}
\begin{center}
\begin{tabular}{|c|cc|cc|}
\hline
Bin & $\cos{(\theta_{\rm obs,max})}$ & $\cos{(\theta_{\rm obs,min})}$ & $\theta_{\rm obs,min}$ (degrees) & $\theta_{\rm obs,max}$ (degrees) \\
\hline
1 & 0.9 & 1.0 & 0.00   &     25.84 \\
2 & 0.8 & 0.9 & 25.84  &     36.87 \\
3 & 0.7 & 0.8 & 36.87  &     45.57 \\
4 & 0.6 & 0.7 & 45.57  &     53.13 \\
5 & 0.5 & 0.6 & 53.13  &     60.00 \\
6 & 0.4 & 0.5 & 60.00  &     66.42 \\
7 & 0.3 & 0.4 & 66.42  &     72.54 \\
8 & 0.2 & 0.3 & 72.54  &     78.46 \\
9 & 0.1 & 0.2 & 78.46  &     84.26 \\
10 & 0.0 & 0.1 & 84.26  &     90.00 \\
\hline
\end{tabular}
\end{center}
\end{table}

Since we have assumed that the torus is comprised of cold matter, photons are always down-shifted in energy in our
Monte-Carlo code. Thus, Green's functions must be calculated for incident energies that are higher than the
highest energy that we require for the final spectra (see \S\ref{evalidity}).  We use an upper incident photon
energy of 500~keV, but this can easily be extended in future work.  Below 5 keV, we adopted a different approach
to explicitly calculate the Green's functions because larger numbers of photons and input energies (and,
therefore, longer run-times) are needed as absorption becomes more important.  The procedure adopted for low
energies is described in \S\ref{loweapprox}.  

We have generated Green's functions for 63 incident energies in the range of $5-500$ keV. Interpolation of these
functions across different incident energies is problematic since there are sharp discontinuities at the
boundaries of the energy distributions for each scattering.  However, the shift in wavelength is linear and the
maximum wavelength shift for a photon with a given incident energy that has been scattered $n$ times is $2n$
Compton wavelengths (see \S\ref{csscater}).  Therefore, the Green's functions have been stored in uniform bins in
units of fractional Compton wavelength shift (\dlcp) to facilitate interpolation since the discontinuities occur
at the same place in these units regardless of incident energy.  The value of \dlc is calculated from the energy
of the escaping photon ($E_{\rm esc}$) and that of the initial injected photon ($E_{\rm in}$) or, in the case of
the emission lines, with respect to the zeroth order, rest-frame line energy: 

\begin{eqnarray}
\Delta\lambda/\lambda_{\rm C} & = & m_{\rm e}c^{2}(\frac{1}{E_{\rm esc}}-\frac{1}{E_{\rm in}}) 
\end{eqnarray}

The total number of Compton wavelengths included in each Green's function is dictated by the largest number of
scatterings experienced by any photon in a given Monte-Carlo run.  For the parameter ranges considered in the
present work, we did not need to account for more than 46 scatterings in the continuum.  We set up 100 uniform
bins per Compton wavelength; this corresponds, for example, to an energy resolution of $\sim 0.8$ eV at 6.4 keV
and $\sim 140$ eV at 100 keV.  At energies above $\sim 15$~keV the limited energy resolution of current -- and
likely near-future -- detectors allows us to use much coarser bin sizes. In the $\sim 5.5-7.5$~keV {\it output}
band we wish to treat the Fe K$\alpha$ and K$\beta$ line profiles with small enough bin sizes to be useful for
future high-spectral resolution ($\sim 1$~eV) detectors. This does not require a fine mesh on the incident photon
array because the line photons are generated in the medium at specific energies. Obtaining a fine resolution line
profile requires only a sufficient number of photons in the line (and therefore in the incident spectrum), so
that the photons can be distributed over a large number of output energy bins with sufficient statistical
accuracy in each bin. Above 10~keV, for the majority of the Green's functions, we have found that $10^{6}$ input
photons are sufficient for the energy resolution we have chosen.  However, at lower energies, especially near the
Fe~K edge, the Monte-Carlo runs have been performed with $10^{7}$ photons, in order to improve the statistical
accuracy in the finer energy bins.  

\subsection{Low-Albedo, Low-Energy Regime}
\label{loweapprox}

As the energy of the photons decreases, the absorption probability becomes increasingly higher and so the
fraction of escaping photons decreases significantly.  In order to obtain reliable escape fractions, larger
input photon numbers must be injected into the torus at low energies.  Moreover, the multitude of
absorption edges below 10 keV would require carefully selected injection energies in order to achieve the
desired energy resolution with sufficient accuracy accuracy (see \S\ref{gridintervals}). However, it is not,
in fact, necessary to calculate the Green's functions on such a fine energy grid when the energy is low
enough that Compton scattering is in the Thomson regime.  In that case, the total scattering cross-section is
essentially independent of energy and, if absorption dominates over scattering, most of the escaping photons
will be zeroth-order or once-scattered.  Under these circumstances, the passage of photons through the torus
depends only on the single-scattering albedo and not the initial energy.  The (energy-dependent)
single-scattering albedo is defined as the scattering cross-section as a fraction of the the total
(absorption plus scattering) cross-section: $s\equiv\sigma_{\rm s}/(\sigma_{\rm a}+\sigma_{\rm s})$.  

Using our Monte Carlo code, we can calculate the number of escaping photons for a relatively small
set of albedo values and interpolate for any desired, arbitrary value of the albedo (see
\fignescvsalbp).  Thus, to obtain the output for a given input photon energy, we can calculate the
corresponding albedo for a given set of element absorption cross-sections at that energy and use
that albedo value to obtain the escape fraction by interpolating the albedo-based Monte Carlo
results.  We note that if any of the element abundances is changed, the same albedo-based
Monte-Carlo results can be used since only the correspondence between albedo and energy changes.  In
practice, we can calculate the zeroth-order photon numbers using Eq. \ref{eq:nzeroth} and use the
interpolated Monte-Carlo results for the scattered photons.  We can then use the numbers of escaping
of photons in one of two ways.  If the energy resolution involved in the application is such that
the energy shifts are negligible, then we can simply place the zeroth-order and scattered photons in
the same energy bin.  At 5 keV the maximum energy shift after one scattering is $\sim100$ eV, which
is comparable to CCD-resolution.  If the energy shifts cannot be neglected, we can use the Thomson
differential cross-section to distribute the once-scattered photons over the correct energy bins.

We utilized Monte-Carlo runs for 17 values of $s$ between $10^{-4}$ and 1 for all of the grid values
of \nh that were used in the full Monte-Carlo simulations (see \S\ref{gridintervals}), each starting
with $10^{7}$ input photons.  There is some overlap between the results from the albedo-based
Monte-Carlo escape fractions and those from the Green's functions that we calculated for $E_{\rm
in}$  down to 5 keV, which allows confirmation of the self-consistency of the two methods (see \S\ref{bbspec}).  

\begin{figure}
\centerline{
	\psfig{figure=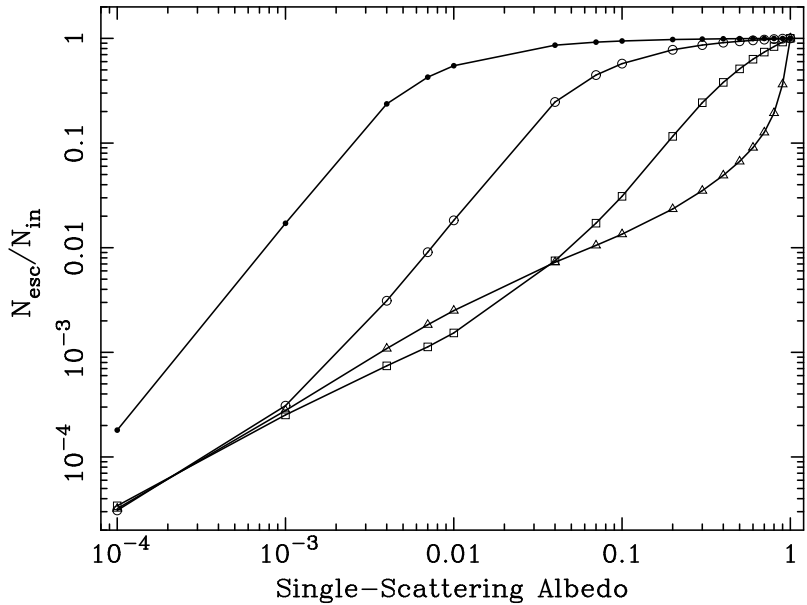,height=8cm}}
\caption{\footnotesize The fraction of the number of escaping photons (including zeroth-order) relative
 to the number of input photons, versus the single-scattering albedo 
 for $N_{\rm H}= 10^{22} \rm \ cm^{-2} \ ({\it filled \ circles}),\ 10^{23} \ cm^{-2} \ ({\it open
\ circles}),\ 
 10^{24} \ cm^{-2} 
\ ({\it squares}),\ and \ 10^{25} \ cm^{-2}$ ({\it triangles}), summed over all \thetaobs bins.
These calculations are for the elastic-scattering regime, and do not depend on the 
input photon energy (see text
for details). }
\end{figure}

\subsection{Energy Range of Validity}
\label{evalidity}

The upper energy range of validity for the integrated spectra calculated from the Green's functions
depends on \nh and \thetaobsp, as well as on the shape of the input spectrum.  At any particular
energy value, the integrated spectrum includes contributions from Compton-down-scattered photons
that had higher initial energies.  In principle, an infinite number of scatterings could contribute
to the total spectrum at a given energy.  In practice, the intrinsic spectrum of an astrophysical
source and the Klein-Nishina cross-section decrease with increasing energy.  Therefore, only a
finite number of scatterings contribute a significant fraction of the photons to the spectrum.  We
can quantify this by, for example, examining the number of scatterings that contribute to 90\% or
more of a given Green's function.  We calculated this number of scatterings as a function of \nh and
\thetaobs from our Monte-Carlo results for the highest input energy (\ein$=500$ keV).  This number,
$n_{\rm max}$, increases from 1 to 6 as \nh increases from $10^{22} \rm \ cm^{-2} \ to \ 10^{25} \rm
\ cm^{-2}$ for all \thetaobsp.  For an initial energy of 500 keV, Compton down-scattering can reduce
the output energy to $E_{f}=511[(511/500)+2n_{\rm max}]^{-1}$ keV.  Since 500 keV is the largest value
of \ein in our grids, this represents the upper limit of the range of validity of models based on
these Green's functions {\it if} the input spectrum extends beyond 500 keV.  \figerange shows this
upper limit as a function of \nh for two values of \thetaobs (face-on and edge-on bins; see
\tablecosrange).  We see that, for \nh$=10^{24} \rm \ cm^{-2}$, this upper limit corresponds to
$\sim169$ keV; for the largest column density (\nh$=10^{25} \rm \ cm^{-2}$), the upper limit is only
$\sim 39$ keV.  If the high-energy cut-off of the intrinsic spectrum of the astrophysical source
being modeled is smaller or equal to 500 keV, the model is valid up to the cut-off energy.  The range
of high-energy cut-offs for the input spectra of AGN is poorly determined, but we note that
non-blazar AGNs have rarely been detected above 500 keV (e.g. see Dadina 2008).

\begin{figure} 
\centerline{\psfig{figure=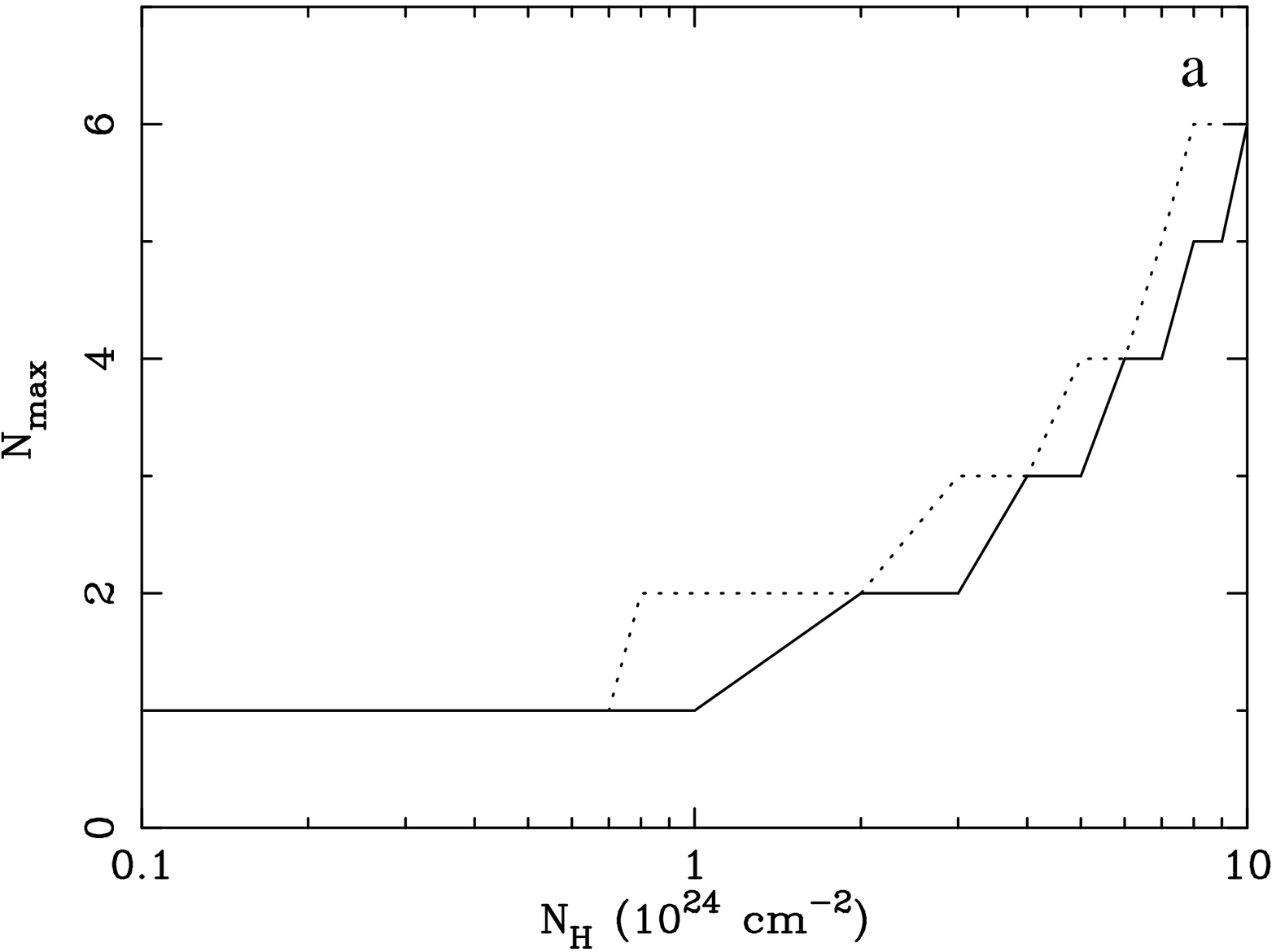,height=7cm}} 
\centerline{\psfig{figure=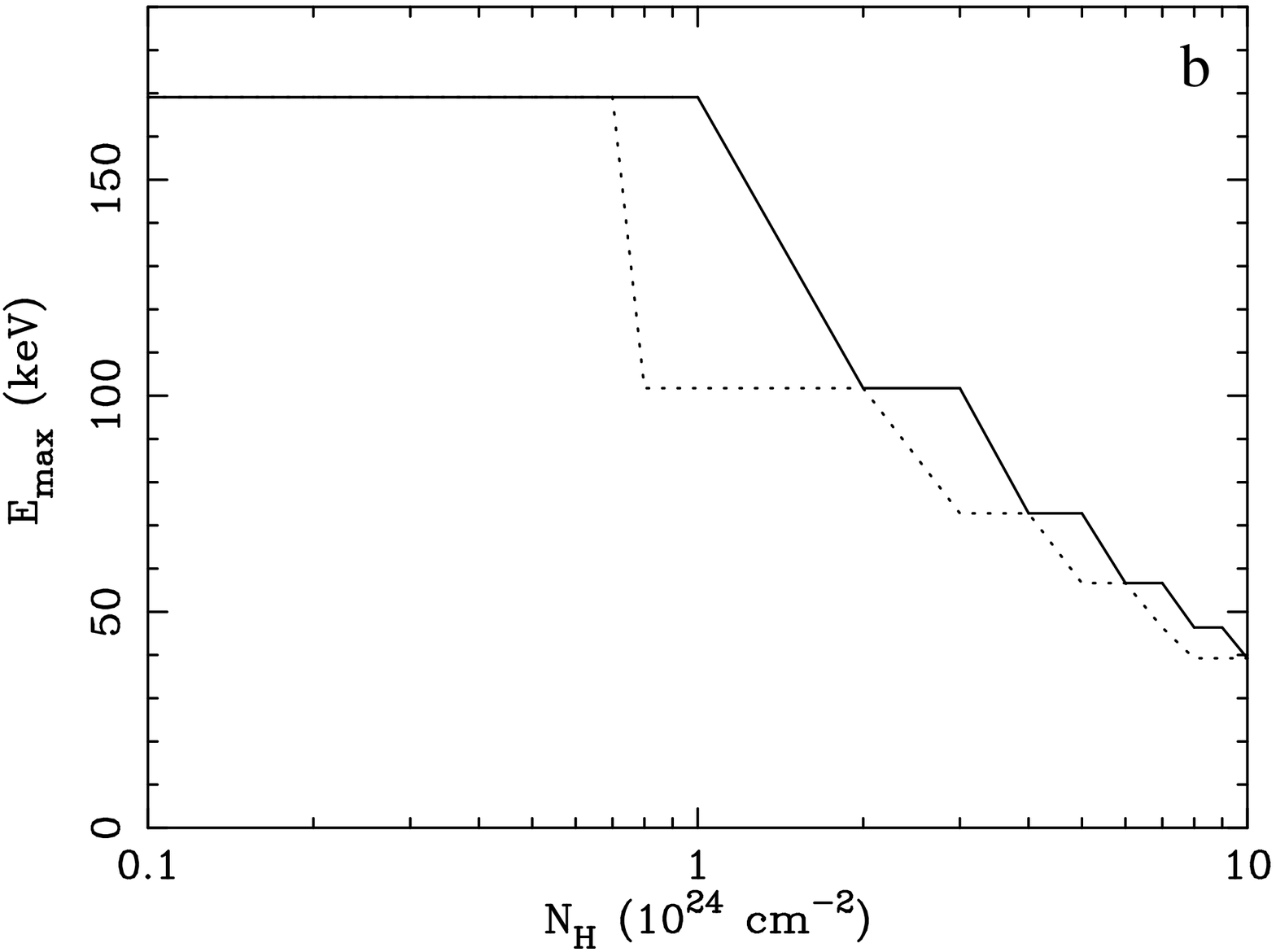,height=7cm}} 
\caption{\footnotesize (a) Maximum number of scatterings experienced by 90\% of the 500~keV 
input photons, 
as described in \S\ref{evalidity}, versus \nhp, for the 
face-on (bin~1 -- {\it dotted curve}) and
edge-on (bin~10 --{\it solid curve}) \thetaobs bins (see \tablecosrange for the \thetaobs bin ranges).
(b) Lowest final energy of 90\% of the 500~keV input photons, versus \nhp, for the 
face-on ({\it dotted curve}) and
edge-on ({\it solid curve}) \thetaobs bins (see \tablecosrange for the \thetaobs bin ranges).} 
\end{figure}

\section{Integrated Spectra}
\label{bbspec}

In the following, we discuss the results of integrating the Green's functions described in \S\ref{greens}, using a
simple power-law input spectrum (flux $\propto E^{-\Gamma} \ \rm keV^{-1}$).  Typical values of the photon index
($\Gamma$) of the intrinsic power-law for the majority of AGN range from 1.5 to 1.9 (e.g. Winter \etal 2008 and
references therein).  We present here examples using $\Gamma=1.5$ and 1.9 to represent this range.  However, we
note that one class of type~1 AGNs, NLS1s (narrow-line Seyfert 1 galaxies), tend to have steeper slopes, up to
$\sim 2.5$.  We do not assume an exponential cut-off above $\sim 100$ keV, as is often assumed in the literature. 
The functional form of a power-law with an exponential cut-off is not physical; in particular, thermal
comptonization models do not predict curvature that can be modeled adequately with an exponential cut-off.  More
importantly, it is useful to obtain broad-band spectral results as a function of the empirical power-law slope
without the extra complication of a cut-off energy.  Then, if necessary, we can infer the results for an input
spectrum with a high-energy cut-off.  Furthermore, Compton-thick reprocessor models {\it themselves}, such as the
one discussed here, introduce curvature in the high-energy spectra and therefore utilizing a simple power-law
input spectrum allows us to assess this intrinsic curvature.

The broad-band spectrum for a given \nh and \thetaobs is created by first setting up an arbitrary energy grid for
the integrated spectrum and, if the line-of-sight intercepts the torus, calculating the zeroth-order continuum
for the given \thetaobs (see Eq. \ref{eq:losnh} and Eq. \ref{eq:nzeroth}).  Then the Green's functions must be
interpolated across the $E_{\rm in}$ table values for each point on the chosen energy grid.  The interpolated
Green's functions are summed and convolved with the chosen input spectrum to produce the scattered spectrum (see
Eq. \ref{eq:greensfn}), which can then be added to the zeroth-order continuum if the line-of-sight intercepts the
torus, or to the intrinsic continuum if the line-of-sight does not intercept the torus.    

The scattered contribution to the fluorescent lines is calculated in a similar manner to that of the
scattered continuum.  For the zeroth-order contribution to the fluorescent lines, the procedure is
similar, but these photons are assigned to a single output energy bin, unique to each fluorescent
line.  

As described in \S\ref{loweapprox}, we have calculated a set of escape fractions for the low-energy, low-albedo
regime, as a function of albedo ($s$) instead of \einp.  The integrated spectra presented here were calculated by
simply assuming elastic scattering in this regime (the small energy shifts may be neglected for the present
purposes -- see \S\ref{loweapprox}).  In this regime, each energy value on the chosen output grid must first be
mapped to an albedo value.  The escape fractions are then interpolated in albedo space, and then translated back
to energy space and multiplied by the input spectrum.  In \figcmpalbedop, we compare the total, integrated
spectrum derived from the primary, full, Monte-Carlo runs (dotted curve) with those obtained from the albedo-based
escape fractions (solid curve) for the same input spectrum (a simple power-law with $\Gamma=1.9$) for \nh$=10^{24}
\rm \ cm^{-2}$ for the case of the edge-on angle bin (bin~10; see \tablecosrangep).  We see, in  general, good
agreement between the two sets of spectra up to $\sim10$ keV.  In all of the examples presented here we choose 5.5
keV as the cross-over energy between the albedo-based spectrum and the spectrum calculated from the primary, full,
Green's functions for the integrated spectra.  We also point out that some of the structure around the Oxygen K
edge seen in the albedo-based integrated spectra below $\sim 0.6$ keV for the edge-on \thetaobs bin is not
physical, but a result of insufficient statistics.    

\begin{figure}
\centerline{
	\psfig{figure=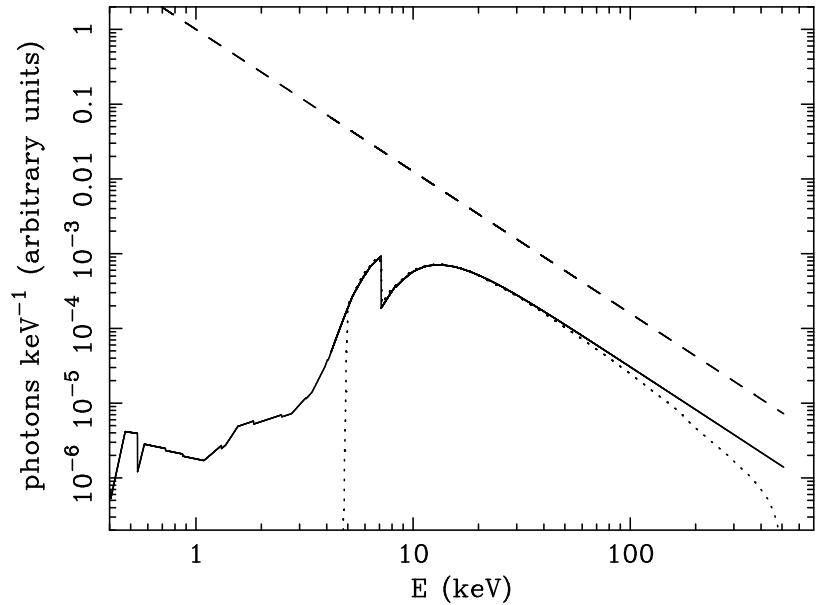,height=8cm}}
\caption{\footnotesize Integrated scattered spectrum from the albedo-based Monte Carlo runs ({\it solid curve}), 
overlaid with the integrated scattered 
spectrum ({\it dotted curve}) from the full Monte-Carlo routine,
for \nh$=10^{24} \ \rm cm^{-2}$ for the edge-on \thetaobs bin (bin~10 -- see \tablecosrangep) 
for an input power-law continuum 
with $\Gamma=1.9$ ({\it dashed line}).}
\end{figure}

\subsection{The Reprocessed Continuum}

We have produced broad-band reprocessed spectra (including the \fekap, \fekbp, and \nika
zeroth-order fluorescent lines and their Compton shoulders) for all of the grid values of \nh and
\thetaobs for $\Gamma=1.5$ and 1.9 (a total of $28 \times 10 \times 2$ reprocessed spectra).  Some
of these spectra are illustrated in \figexspecfigsp, which shows spectra for the face-on \thetaobs
bin (bin~1 -- {\it dotted curves}; see \tablecosrangep) and  edge-on \thetaobs bin (bin~10 -- {\it
solid curves}; see \tablecosrangep) for \nh$=5\times10^{23} \rm \ cm^{-2} \ ({\it top \ left}), \
10^{24} \rm \ cm^{-2} \ ({\it top \ right}), \ 5\times10^{24} \rm \ cm^{-2} \  ({\it bottom \
left}), \ and \ 10^{25} \rm \ cm^{-2} \ ({\it bottom \ right})$ for an input power-law continuum 
with $\Gamma=1.9$ ({\it dashed line}).   The zeroth-order spectrum for the edge-on \thetaobs bin is
also shown ({\it dot-dashed curves}) for each \nhp.  Note again that, for the edge-on angle bin, the
spectrum below $\sim0.6$ keV is subject to significant statistical error and should be interpreted
with caution.  For the five angle bins that intercept the torus, (\thetaobs$\ge60^{\circ}$ -- see
\tablecosrangep), the reprocessed spectrum includes the zeroth-order (transmitted) continuum as well
as the scattered continuum.  For the five non-intercepting angle bins (\thetaobs$<60^{\circ}$), the
reprocessed spectrum includes only the scattered continuum.  Therefore, in order to obtain the {\it
total} observed spectrum in the non-intercepting angle bins, we must add the intrinsic continuum to
the reprocessed spectra.  It should be noted that in \figexspecfigs and in subsequent figures {\it
we show only the reprocessed spectrum} for the non-intercepting angles in order to directly compare
spectral features for face-on and edge-on angle bins.  From \figexspecfigs we see that the Compton
hump, as expected, begins to appear for column densities of $10^{24} \rm \ cm^{-2}$ and higher and
is stronger for the edge-on (solid) spectra since the photons that are scattered into the edge-on
angle bin in general pass through a larger Thomson depth than those that are scattered into the
face-on angle bin (dotted).  At low energies, although the continuum is diminished, we do not see a
complete extinction of the spectrum.  In contrast, a pure-absorption model or a spherical
Compton-thick model would predict a stronger suppression of the low-energy photons (e.g. Leahy \&
Creighton 1993; Yaqoob 1997), especially for column densities of $10^{24} \rm \ cm^{-2}$ and
higher.  Our torus model instead results in a non-negligible number of photons being scattered into
the line-of-sight at low energies, even for the edge-on observing angle of a high-\nh torus, due to
the smaller line-of-sight column densities towards the surface of the torus.  At high energies, we
find that the spectra for the face-on angle bin cut off at lower energies than those for the edge-on
angle bin.  This is due to the diluting effect of the zeroth-order contribution to the reprocessed
spectra in the angle bins that intercept the torus.  For \nh less than $\sim 5 \times 10^{24} \rm \
cm^{-2}$, the edge-on reprocessed spectra are dominated by the zeroth-order photons. However, we
find that, for high \nh values (greater than $\sim 5\times10^{24} \rm \ cm^{-2}$) even the edge-on
high-energy spectra are suppressed (since the zeroth-order photons no longer dominate).

To illustrate the difference in the relative amplitude of the scattered spectra as a function of
observing angle, we show in \figscatvstheta the ratio of the scattered-to-intrinsic continuum at a
single energy (6.4~keV) for an input power-law continuum with $\Gamma=1.9$, versus \thetaobsp. 
These ratios are shown for \nh$=5\times10^{23} \rm \ cm^{-2} \ ({\it dot-dashed \ curve}), \
8\times10^{23} \rm \ cm^{-2} \ ({\it dashed \ curve}), \ 2\times10^{24} \rm \ cm^{-2} \  ({\it
dotted \ curve}), \ and \ 10^{25} \rm \ cm^{-2} \ ({\it solid \ curve})$.  For all values of \nhp,
the ratios show little dependence on the inclination angle for values of \thetaobs that do not
intercept the torus.  In the Thomson-thin regime, as would be expected, there is little distinction
in the ratio of the reprocessed-to-intrinsic continuum between intercepting and non-intercepting
lines of sight, in contrast with the behavior for larger column densities.  

\begin{figure}
\centerline{
\psfig{figure=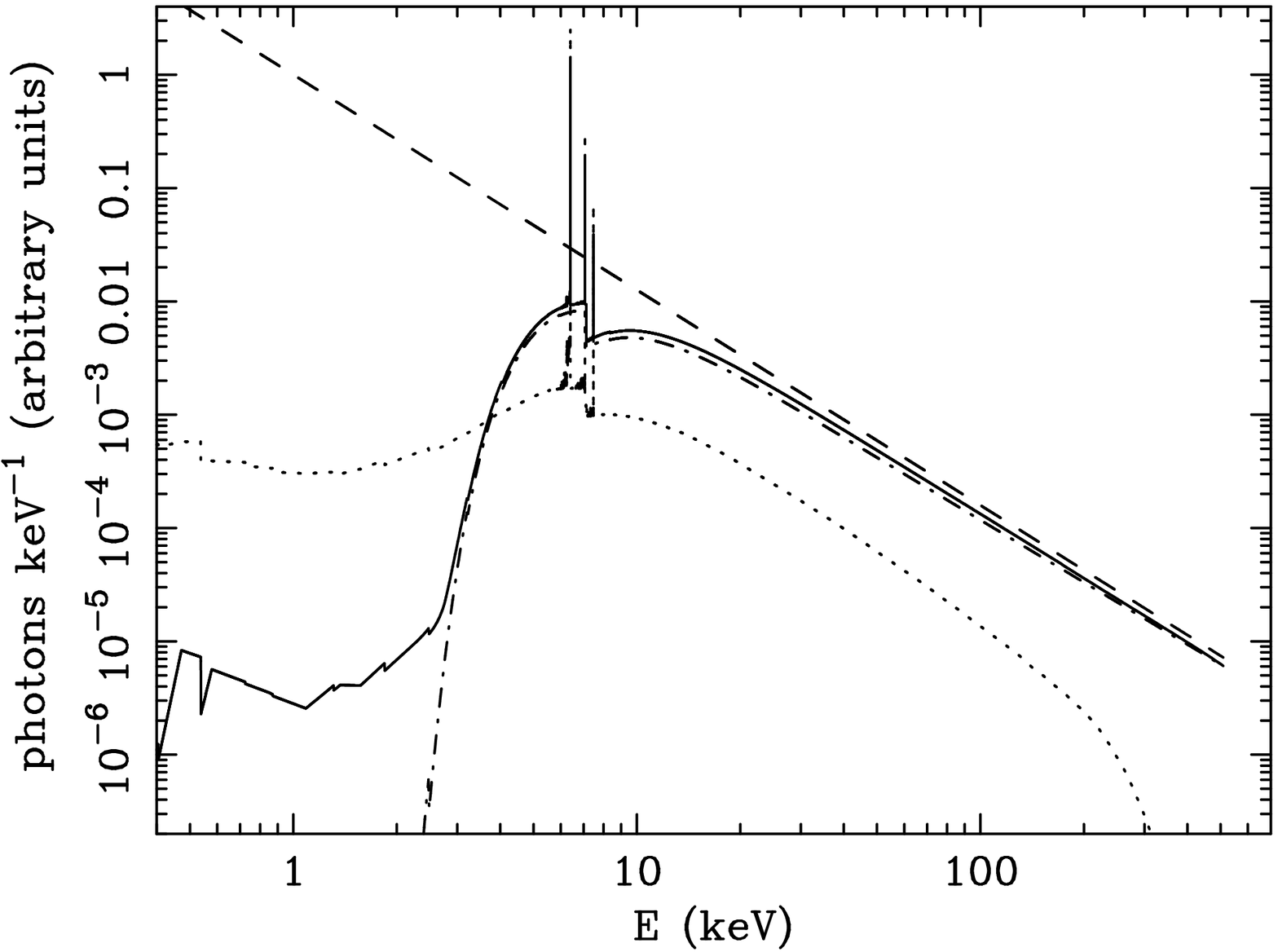,height=5.5cm}
\psfig{figure=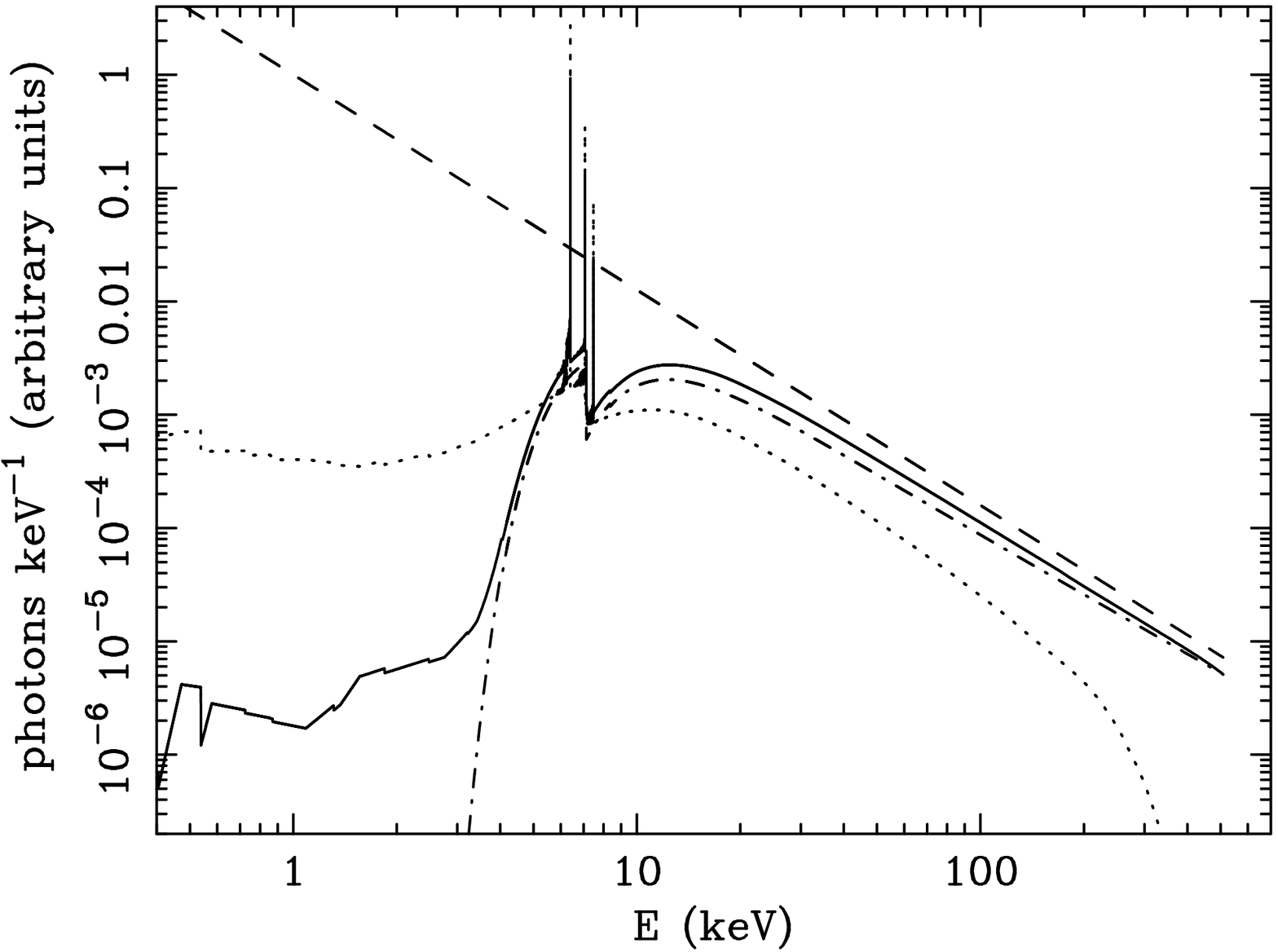,height=5.5cm}}
\centerline{
\psfig{figure=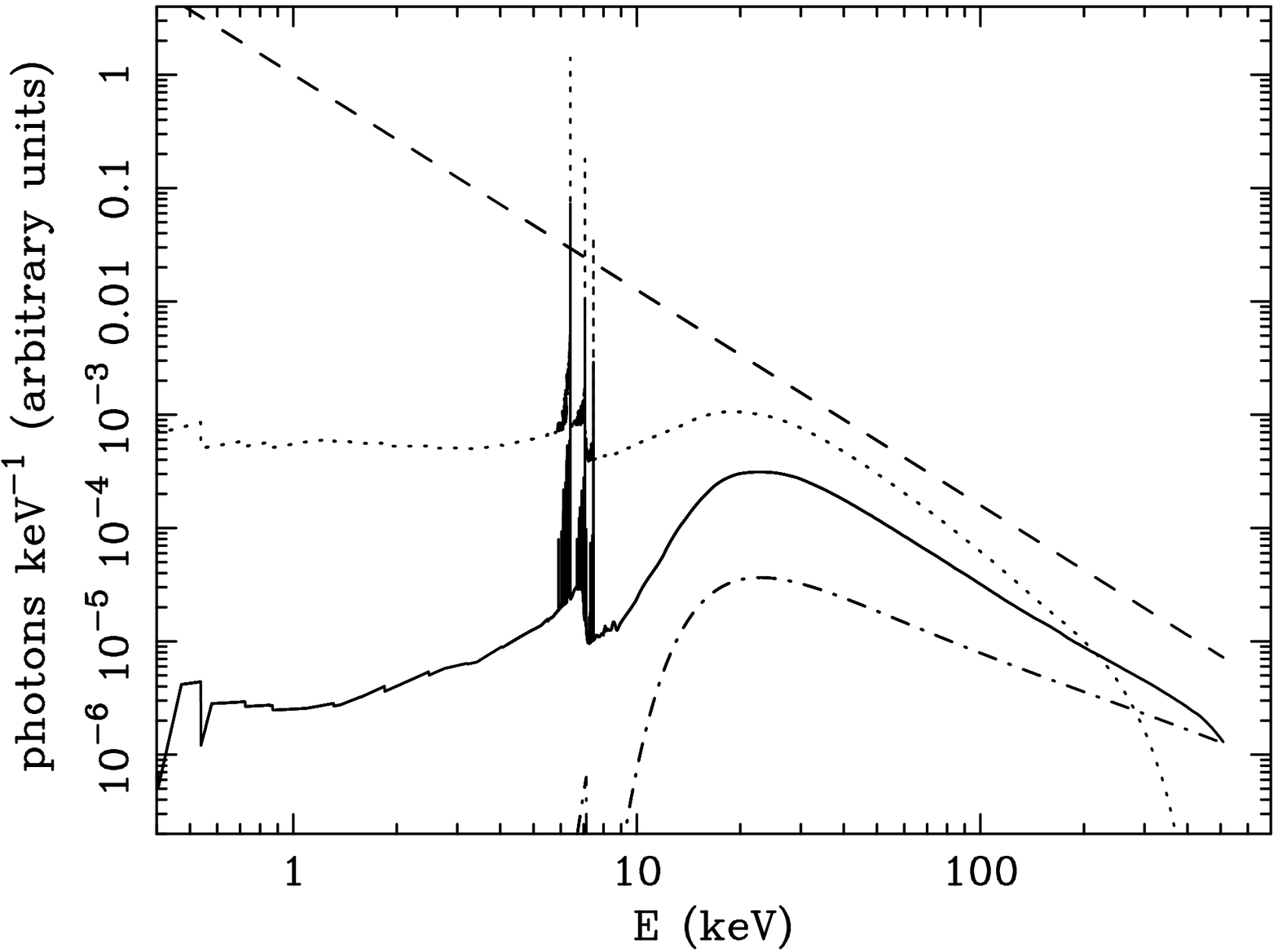,height=5.5cm}
\psfig{figure=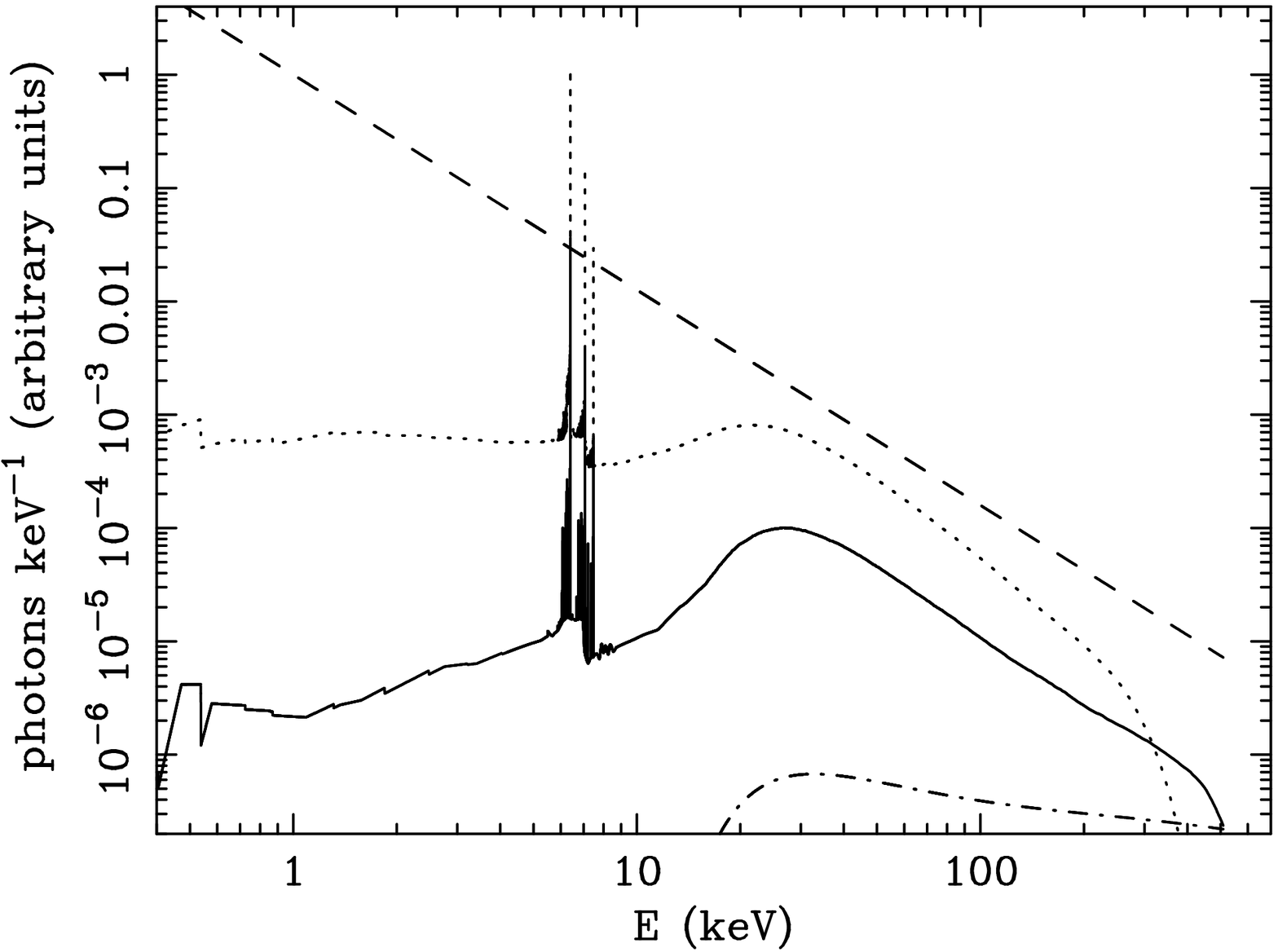,height=5.5cm}}
\caption{\footnotesize Total integrated spectra for an input power-law continuum 
with $\Gamma=1.9$ ({\it dashed
line}). Spectra for the face-on \thetaobs bin (bin~1 -- {\it dotted  curves}; see \tablecosrangep) and 
edge-on \thetaobs bin (bin~10 -- {\it solid curves}; see \tablecosrangep) are shown 
for \nh$=5\times10^{23} \rm \ cm^{-2} \ ({\it top \ left}),
\ 10^{24} \rm \ cm^{-2} \ ({\it top \ right}), \ 5\times10^{24} \rm \ cm^{-2} \ 
({\it bottom \ left}), \ and \
10^{25} \rm \ cm^{-2} \ ({\it bottom \ right})$.  The zeroth-order spectrum for the edge-on
\thetaobs bin is shown ({\it dot-dashed curves}) for each \nhp.  Note that, for the edge-on angle
bin, the spectrum below $\sim0.6$ keV is subject to significant statistical error and should be
interpreted with caution.}
\end{figure}

\begin{figure}
\centerline{
	\psfig{figure=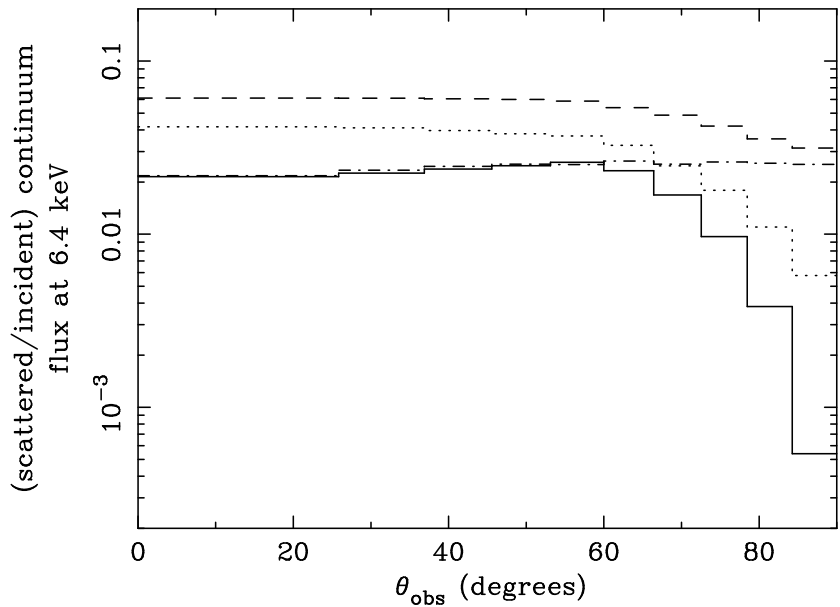,height=8cm}}
\caption{\footnotesize The ratio of the number of scattered photons at 6.4~keV that have escaped from 
the torus 
to the incident continuum photons at 6.4~keV, versus \thetaobs for \nh$=5\times10^{23}
 \rm \ cm^{-2} \ ({\it dot-dashed \ curve}),
\ 8\times10^{23} \rm \ cm^{-2} \ ({\it dashed \ curve}), \ 2\times10^{24} \rm \ cm^{-2} \ 
({\it dotted \ curve}), \ and \
10^{25} \rm \ cm^{-2} \ ({\it solid \ curve})$.}
\end{figure}

Thus far, we have focused on the reprocessed spectra that result from a power-law input spectrum with
$\Gamma=1.9$.  We have confirmed that the chosen value of the spectral index has little effect on the general
behavior of the integrated, reprocessed continuum.  However, we note that for the scattered continuum, the
amplitude of the Compton hump relative to the spectral flux below 10~keV is larger for flatter incident
continua. We find that the maximum ratio of the scattered to incident  flux (i.e., at the peak of the Compton
hump) for $\Gamma=1.5$ is $\sim 25$\% higher than the same ratio for $\Gamma=1.9$ for $N_{H} = 10^{25} \  \rm
cm^{-2}$ and the face-on angle bin (bin~1, \tablecosrangep). For the edge-on angle bin and the same column
density, the corresponding enhancement of the Compton hump for $\Gamma=1.5$ relative to $\Gamma=1.9$ is $\sim
35$\%.

\subsubsection{Comparison with {\tt PEXRAV}}
\label{pexrav}

It is common practice is to use a disk-reflection model (in particular, {\tt pexrav} in XSPEC --
see Magdziarz \& Zdziarski 1995) to attempt to imitate scattering in the torus or other non-disk
geometry.  The disk geometry is clearly incongruent to the geometry of the distant-matter
reprocessor and therefore the derived parameters are not relevant or physical.  It is useful to
directly compare the reprocessed spectra from the angle bins that do not intercept the torus with
spectra obtained with {\tt pexrav} in order to assess the effect on spectral-fitting results reported in the
literature.  The key parameters of the disk-reflection model are the inclination angle of the disk
normal relative to the observer's line-of-sight, and the so-called ``reflection fraction'', $R$.
The latter is a multiplication factor for the reflected spectrum normalization obtained from a
semi-infinite, neutral disk illuminated by a non-varying X-ray  point-source continuum. Thus, if the
disk-reflection model is applied to a situation for which the disk geometry is not appropriate, the
value of $R$ that is obtained to force a fit to data no longer has a meaningful interpretation.
Nevertheless, the results of forced disk-reflection fits to obscured AGN abound in the literature. 

\figcmppexrav shows a comparison of an integrated spectrum obtained from our Monte-Carlo torus
results (for an input spectrum with $\Gamma=1.9$  and \nh$=10^{25} \rm \ cm^{-2}$) with spectra
obtained with the {\tt pexrav} model in XSPEC.  The torus spectrum is shown for the face-on \thetaobs
bin (bin~1; dotted curve).  Overlaid are the {\tt pexrav} results for the smallest value of \thetaobs
allowed by that model ($18^{\circ}$) with two reflection fraction values ($R=1$ and $R\sim0.16$) as
well as for \thetaobs$=60^{\circ}$ with $R=1$.  It can be seen that, relative to the intrinsic
power-law, the reflection continuum from our torus model is significantly weaker than both of the
{\tt pexrav} components with $R=1$.  The value of $R\sim0.16$ simply aligns the face-on, $R=1$ {\tt
pexrav} component approximately with the torus component at the position of the Fe~K edge.  However,
the detailed shape of the soft X-ray continuum and the Compton hump from the torus model are
different compared to the {\tt pexrav} spectra.  These differences are due to geometry effects and
are significant enough to potentially impact fitting results for high signal-to-noise data.  Even for
lower signal-to-noise data, our results show that, for the angle bin that gives the largest
reflection spectrum from the torus (i.e., face-on), the reflected continuum is suppressed by a factor
$\sim 6$ compared to a disk geometry. 

The difference between the Compton-thick reflection spectrum from our torus model and that from a
disk is {\it substantial}.  This is due to several effects. One is that when the torus becomes
Compton-thick, half of the photons reflected from its surface are directed away from the observer.
The projection effects due to the curved toroidal surface, and re-entry of some of the reflected
photons into the torus also play a role. The fact that face-on viewing angles in the adopted geometry
disfavor forward and back-scattered photons contributes to angle-dependent reflection via the
differential scattering cross-section. A more significant factor is that the amplitude of the
reflection spectrum depends critically  on the angle of reflection relative to the local normal on
the toroidal surface. As this angle varies over the toroidal surface between $90^{\circ}$ and the
complement of the half-opening angle  (i.e., $30^{\circ}$ for a half-opening angle of $60^{\circ}$)
for the face-on orientation, the reflection spectrum amplitude varies from zero to some value that is
less than that for a $0^{\circ}$ reflection angle. The net reflection spectrum amplitude will then be a
function of the range in reflection amplitude integrated over all reflection angles over the toroidal
surface. Thus, the virtually universal practice of interpreting the so-called ``reflection-fraction''
as a literal covering factor, or solid angle, from fitting disk-reflection spectra is erroneous. We
have shown that even though our torus subtends the same solid angle at the X-ray source as a disk
($2\pi$), the reflection spectrum amplitude from the torus is a factor of $\sim 6$ less than that
from the disk. Even different toroidal geometries will give different amplitudes for the
Compton-thick reflection spectra for the same torus half-opening angle. For example the
spherical-torus geometry of Ghisellini \etal (1994) and Ikeda \etal (2009) will give different
reflection continuum  amplitudes because the reflection angle relative to a local normal anywhere on
the reflecting surface is constant. For the particular parameters adopted by Ikeda \etal (2009),  a
half-opening angle of $60^{\circ}$ gives a reflection angle of $\sim 30^{\circ}$ so in that case the
reflection spectrum will be stronger than that in our models.

Changing the opening angle of the torus does not change our conclusion about the weakness of Compton-thick
reflection for our geometry (compared to that for a disk). Increasing the torus open angle decreases the
solid angle subtended at the X-ray source and thus reduces the reflection spectrum. Decreasing the opening angle
of the torus increases the magnitude of the reflection spectrum,  but this is countered by a decrease due
to the increase in the average angle of reflection.  A half-opening angle somewhere between $30^{\circ}$
and $45^{\circ}$ will maximize the amplitude of the reflection spectrum but it will never be greater than
a factor of $2$ relative to the half-opening angle of $30^{\circ}$ (e.g. see Ikeda et al. 2009). Although
a direct  comparison between the Compton-thick reflection spectrum from the non-intercepting angle bins of
our torus model and that from a disk is simple, a comparison for obscured lines-of-sight is not
straight-forward.  For inclination angles that intercept the reprocessor it is not possible to assign a
useful meaning to fitted values of $R$ since the reprocessed spectrum is typically modeled in the
literature by an {\it ad~hoc} combination of pure absorption plus disk-reflection.

\begin{figure}
\centerline{
\psfig{figure=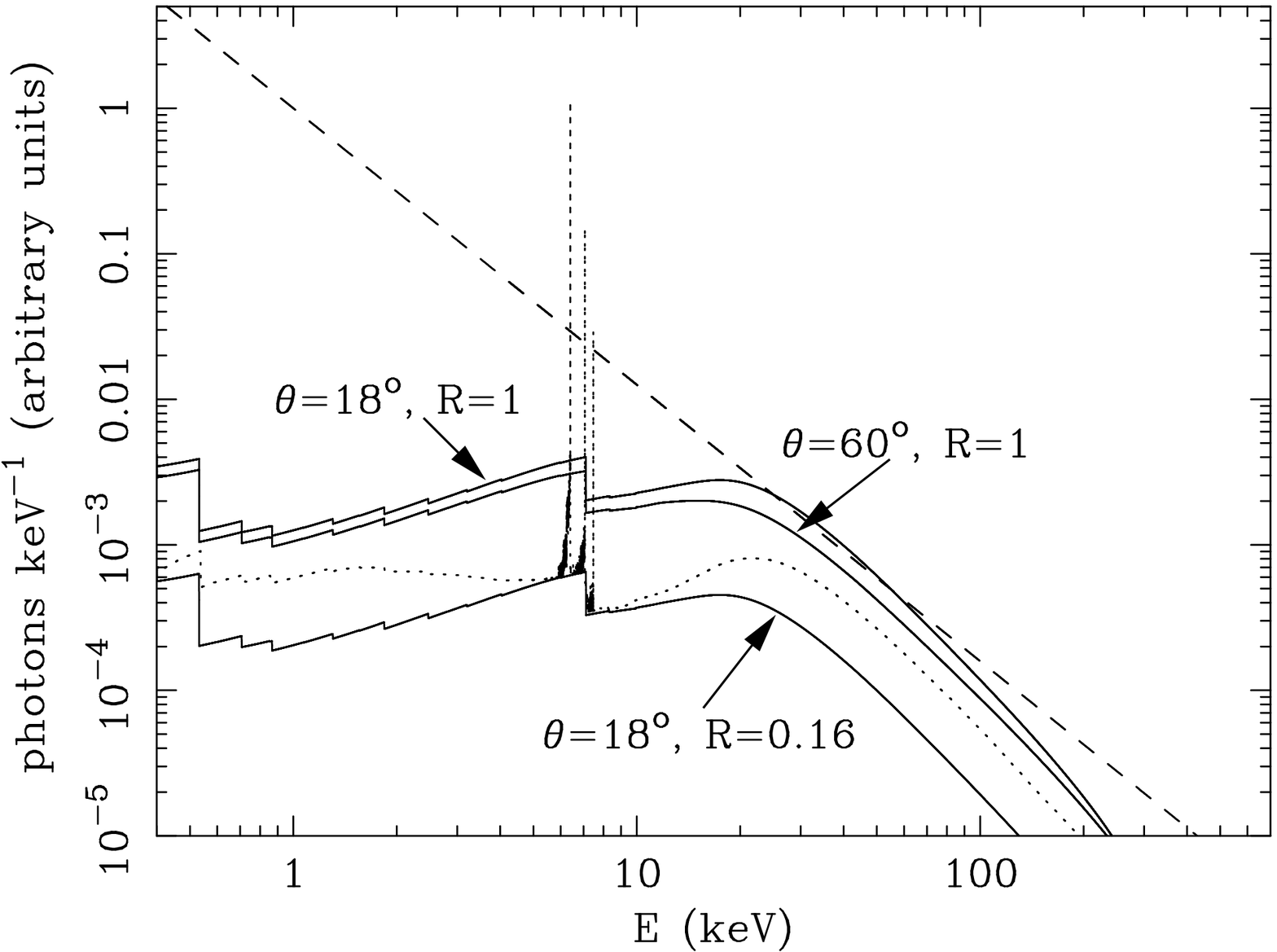,height=8cm}}
\caption{\footnotesize Comparison of an integrated spectrum obtained from our Monte-Carlo torus results 
with those obtained with the {\tt pexrav} model in XSPEC for an input power-law spectrum 
with $\Gamma=1.9$ (dashed line)
and \nh$=10^{25} \rm \ cm^{-2}$.  The torus spectrum is shown for the face-on \thetaobs
bin (bin~1 -- {\it dotted}; see \tablecosrangep).  Overlaid are the {\tt pexrav} results for the 
smallest value of 
\thetaobs allowed by the model ($18^{\circ}$) for two values of the reflection normalization, 
$R=1$ and $R\sim0.16$.  Also shown is the spectrum obtained with {\tt pexrav}, for 
\thetaobs$=60^{\circ}$ and $R=1$. }    
\end{figure}

\subsection{The \feka Emission Line}
\subsubsection{Fluorescent Line Equivalent Widths}
\label{lineew}

The equivalent widths (EWs) of the fluorescent lines are an important diagnostic of the material out
of the line-of-sight.  Observational measurements of the EWs of the fluorescent \feka emission line
in the literature generally refer to the zeroth-order emission only, since the line is typically
fitted with a Gaussian.  The scattered flux (i.e., the Compton shoulder, which can carry up to 40\% of
the zeroth-order flux -- see \S\ref{linecs}) is generally not fitted (usually because it is not
detected), or, if it is fitted, a separate EW is quoted.  Yet these zeroth-order line EWs are often
compared to theoretical predictions that include {\it all} of the line photons (zeroth-order and
scattered; e.g. Leahy \& Creighton 1993, Ghisellini \etal 1994, Ikeda \etal
2009).  In \figewvsnhp, we show the zeroth-order EWs of the \feka line as a function of the column
density of the torus, \nhp, since it is the zeroth-order emission that is observationally-relevant
for the main peak of the line.  The lower curves show the results for the non-intercepting angle
bins, with ascending \thetaobs bins from top to bottom, and the upper curves show the results for the
intercepting angle bins, with ascending \thetaobs bins from bottom to top. 

It can be seen in \figewvsnh that, for \nh less than $\sim 5\times10^{22} \ \rm cm^{-2}$,
inclination-angle effects are not important (as would be expected for the Thomson-thin limit). 
Overlaid on \figewvsnh is the theoretical Thomson-thin limit ({\it dashed line}), given by

\begin{eqnarray}
\label{eq:ew}
EW_{\rm Fe-K\alpha} & = &
785
\ \left(\frac{\Delta\Omega}{4\pi}\right)
\ \left(\frac{\omega_{K}}{0.347}\right)
\ \left(\frac{\rm A_{\rm Fe}}{4.68 \times 10^{-5}}\right)
\ \left(\frac{\sigma_{0}}{3.37\times10^{-20} \rm \ cm^{2}}\right)
\ \left(\frac{3.57}{\Gamma + \alpha -1}\right) 
\ \left(0.8985\right)^{(\Gamma-1.9)}
\ N_{24}\ \rm eV
\end{eqnarray}
(see Yaqoob \etal 2001 for details, but note that Eq. \ref{eq:ew} differs from the corresponding equation
in Yaqoob \etal 2001 since the latter erroneously included the total cross-section at the Fe~K edge
threshold, instead of the K-shell cross-section only).

The quantity $(\Delta\Omega/4\pi)$ is the solid angle of the line-emitting matter subtended at the
X-ray source.  For the torus, $N_{24}=(\pi/4)(N_{\rm H}/10^{24} \rm \ cm^{-2})$ is the column
density averaged over all incident photon angles.  The K-shell fluorescence yield is given by
$\omega_{K}$ and $A_{\rm Fe}$ is the Fe abundance relative to Hydrogen.  The quantity $\sigma_{0}$
is the Fe~K shell absorption cross-section at the Fe~K edge, and $\alpha$ is the power-law index of
the cross-section as a function of energy.  We fitted photoelectric K-shell absorption cross-section
data from Verner \etal (1996) tables with a simple power-law and obtained 
\begin{equation}
\sigma_{\rm Fe-K}(E) = \sigma_{0} 
\left(\frac{E}{E_{K}}\right)^{-2.67} \times 10^{-20} \ \rm cm^{2}
\ per \ atom 
\end{equation}
for $E \ge E_{K}$, giving $\alpha=2.67$ and $\sigma_{0}=3.37$, where $E_{K}=7.124$ keV in Verner \etal
(1996). 

\begin{figure}
\centerline{
\psfig{figure=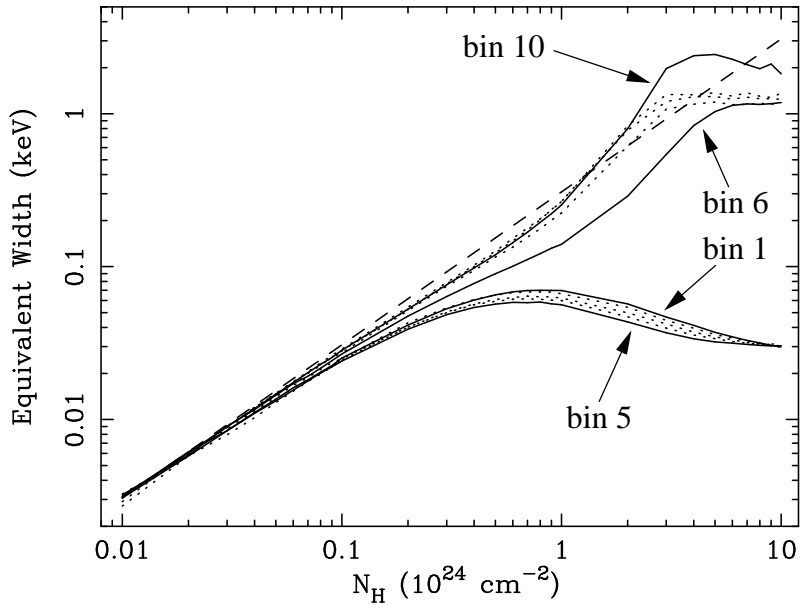,height=8cm}}
\caption{The \feka line equivalent width versus \nh for $\Gamma=1.9$ as calculated from the Monte-Carlo
code, and the theoretical Thomson-thin
limit ({\it dashed line}).  Curves are shown for each of the 10 \thetaobs bins (see \tablecosrangep).  The lower
set of curves corresponds to bins 1--5 (lines-of-sight that do not intercept the torus) and the upper
set of curves correspond to bins 6--10 (lines-of-sight that intercept the torus).  The
{\it solid} curves correspond to the two boundary \thetaobs bins (bins 1, 5, 6, and 10; see \tablecosrangep) 
for both intercepting and non-intercepting cases.  }
\end{figure}

In the limit of small \nhp, the relation between the EW and \nh is linear, for all observing angles,
regardless of the value of $\Gamma$.  The slope of this relation depends only on the element abundance,
$\Gamma$, and $\Delta\Omega/4\pi$.  It can be seen in \figewvsnh that inclination-angle effects begin to
appear even for \nh$\sim10^{23} \rm \ cm^{-2}$.  

For inclination angles that do not intercept the torus, the EW peaks between $\sim5\times10^{23} \rm
\ cm^{-2} \ and \ 10^{24}\ cm^{-2}$, then decreases by $\sim50$\% of this peak value at \nh$=
10^{25}\ \rm cm^{-2}$.  The peak value for the EW of the \feka line, for $\Gamma=1.9$, is $\sim70$ eV
(for the case of $\Gamma=1.5$, the peak value is $\sim 85$ eV).  The decrease in the EWs at the
higher values of \nh is due to the fact that the absolute number of escaping line photons decreases,
partly because more escaping photons appear in the Compton shoulder (see \S\ref{linecs}). 
Additionally, the total continuum is dominated by the intrinsic continuum, which is not affected by
the torus at these inclination angles.  \figewvsnh shows that the limiting \feka line EW for
Compton-thick reflection can be as little as $\sim 30$~eV, a factor of $\sim 5-6$ less than that
expected from a face-on disk (e.g. see George \& Fabian 1991). The reasons for this are the same as
those discussed for the relative weakness of the Compton-thick reflection continuum (\S\ref{pexrav}).
The spherical-torus geometry adopted by Ghisellini \etal (1994) and Ikeda \etal (2009) can yield
larger values of the limiting Fe~$K\alpha$ line EW for Compton-thick reflection for the same
half-opening angle as in our calculations. Again this is due to the same reasons that their geometry
can give a larger reflection continuum, as discussed in \S\ref{pexrav}.

 
\figewvstheta shows the \feka line EW versus \thetaobs for $\Gamma=1.9$  and \nh$=5\times10^{23} \rm
\ cm^{-2} \ ({\it dot-dashed}), \ 8\times10^{23} \rm \ cm^{-2} \ ({\it dashed}), \ 2\times10^{24} \rm
\ cm^{-2} \  ({\it dotted}), \ and \ 10^{25} \rm \ cm^{-2} \ ({\it solid})$.  For the
non-intercepting angle bins the EW is not very sensitive to $\theta_{\rm obs}$, but the face-on EW
can be $\sim 30$\% higher than the EW for the largest non-intercepting value of $\theta_{\rm obs}$
when the continuum at 6.4~keV is dominated by one Compton scattering (i.e., $N_{H} \sim 1-2 \times
10^{24} \ \rm cm^{-2}$). This is because the form of the differential scattering cross-section in the
Thomson regime preferentially produces more forward-scattered and back-scattered photons compared to
those with intermediate scattering angles. In our toroidal geometry this means that there are less
continuum photons scattered into the face-on angle bin than those with larger inclination angles when
only one scattering dominates. Since the zeroth-order \feka line photons are emitted
isotropically, the net result is an enhanced EW for smaller inclination angles in the column-density
regime for which the Thomson depth is of order unity.  For the torus-intercepting angle bins, the EWs
of the lines increase as \nh increases up to $\sim 3\times10^{24} \rm \ cm^{-2}$ since the
zeroth-order continuum, which dominates the continuum at these column densities, decreases due to an
increasing line-of-sight column density (see \figewvsnhp).  Above $\sim 3\times10^{24} \rm \
cm^{-2}$, it is the scattered continuum that begins to dominate the spectrum.  Since both the
absolute flux of the emission lines and that of the scattered continuum effectively decrease at a
similar rate as \nh increases, the EWs of the lines saturate at high \nhp, where the scattered
continuum dominates.  The EWs obtained for the intercepting angle bins are very sensitive to
\thetaobs for \nh greater than $2\times10^{23} \rm \ cm^{-2}$ due to the diminishing zeroth-order
contribution to the continuum; at a given \nh the EW becomes larger as \thetaobs increases (as seen
in \figewvsthetap).  For the edge-on angle bin, the maximum values of the \feka EW are $\sim3.8$ keV
and $\sim2.5$ keV for $\Gamma=1.5$ and 1.9, respectively.   

\begin{figure}
\centerline{
\psfig{figure=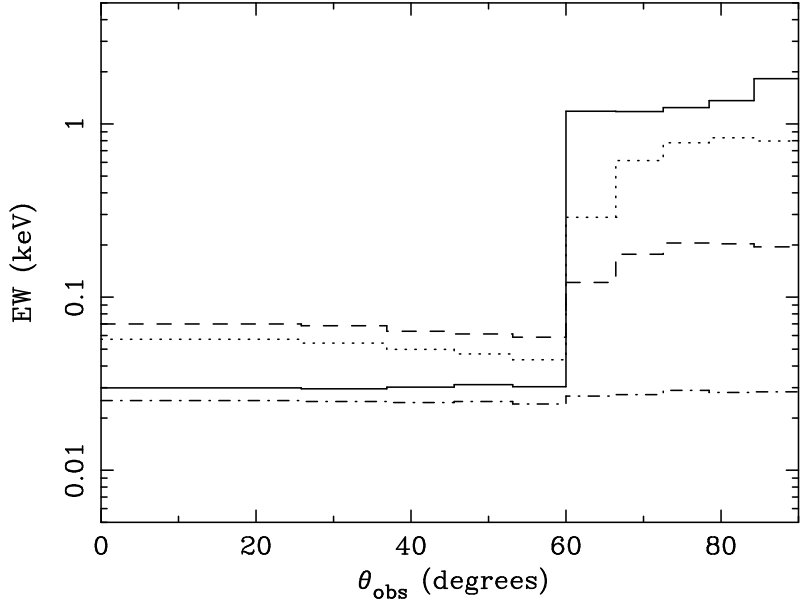,height=8cm}}
\caption{\footnotesize The \feka line equivalent width versus \thetaobs for $\Gamma=1.9$ 
and \nh$=5\times10^{23} \rm \ cm^{-2} \ ({\it dot-dashed}),
\ 8\times10^{23} \rm \ cm^{-2} \ ({\it dashed}), \ 2\times10^{24} \rm \ cm^{-2} \ 
({\it dotted}), \ and \
10^{25} \rm \ cm^{-2} \ ({\it solid})$.}
\end{figure}

We find that smaller values of $\Gamma$ yield larger values of the \feka EW; this is expected as there
are relatively more photons in the continuum above the Fe~K edge for flatter spectra.  \figewratvsnh
shows the ratio of the \feka line equivalent width for $\Gamma=1.5$ to the corresponding equivalent width
for $\Gamma=1.9$, versus \nhp, for each of the inclination-angle bins (see \tablecosrangep).  In the
Compton-thick regime, the ratio of the equivalent widths for the non-intercepting angle bins can be as
large as 1.3 and for the intercepting bins it can be as large as 1.6.

\begin{figure}
\centerline{
\psfig{figure=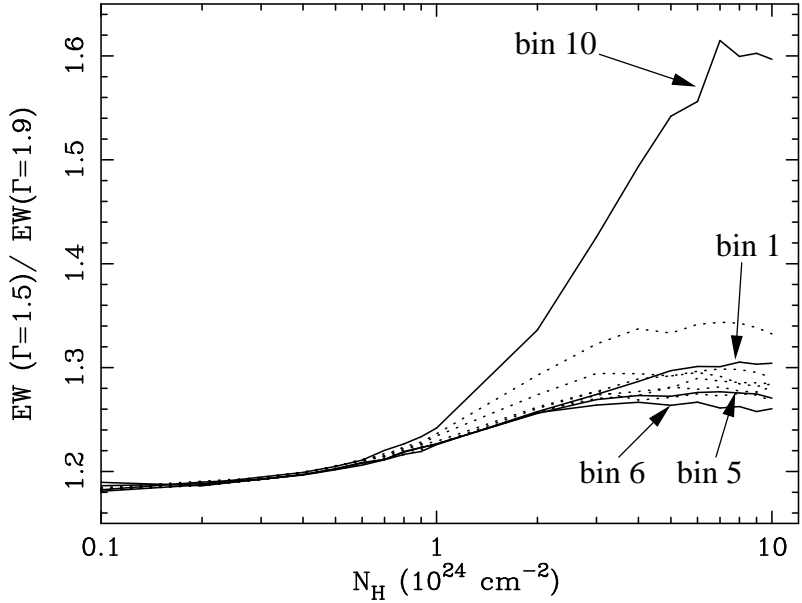,height=8cm}}
\caption{\footnotesize Ratio of the \feka line equivalent width for $\Gamma=1.5$ to the corresponding
equivalent width for $\Gamma=1.9$, versus \nhp.  Curves are shown for each of the 10 
\thetaobs bins (see \tablecosrangep).  The lower
set of curves corresponds to bins 1--5 (lines-of-sight that do not intercept the torus) and the upper
set of curves correspond to bins 6--10 (lines-of-sight that intercept the torus).  The
{\it solid} curves correspond to the two boundary \thetaobs bins (bins 1, 5, 6, and 10; see \tablecosrangep) 
for both intercepting and non-intercepting cases.  }
\end{figure}

The EW versus \nh curves have an explicit dependence on the assumed opening angle of the torus.  This is
because different opening angles correspond to different solid angles subtended by the torus at the source
and to different projection-angle effects.  In the Thomson-thin regime, this dependence is linear (see Eq.
\ref{eq:ew}).  In the Compton-thick regime, there is a more complicated dependence that must be determined
by additional Monte-Carlo simulations and will be the subject of future investigation (see Ikeda \etal
2009 for discussion in the context of a different toroidal geometry).

\subsubsection{\feka Line Compton Shoulder}
\label{linecs}

In addition to the zeroth-order core of the \feka emission line, the shape and relative magnitude of
the Compton shoulder are also sensitive to the properties of the reprocessor.  Evidence for the
Compton shoulder has already been found in some sources (e.g. see Iwasawa, Fabian, \& Matt 1997;
Matt 2002; Kaspi \etal 2002; Watanabe \etal 2003; Yaqoob \etal 2005), although detections of the
Compton shoulder to date have been ambiguous.  Future detectors, such as X-ray calorimeters, will be able
to measure the line  profile structure in more detail than is currently possible.

In this section we discuss the relative strength of the scattered component of the \feka emission
line (i.e., the Compton shoulder) with respect to the zeroth-order line component and the continuum. 
\figcmpcsa shows ratio plots of the total number of scattered \feka line photons to zeroth-order
\feka line photons versus \nhp.  The ratios are shown for an input power-law continuum with
$\Gamma=1.9$, for two \thetaobs bins (face-on -- bin~1 -- {\it dotted curve} and edge-on
-- bin~10 -- {\it solid curve};  see \tablecosrangep).  Note that, for \nh$<10^{23} \rm \
cm^{-2}$, these ratios are  affected by small-number statistics and should be interpreted with
caution.  The general trend is that the ratios increase as \nh increases, up to a column density
between $10^{24} \rm \ cm^{-2} \ and \ 5\times10^{24} \ cm^{-2}$, since the Thomson depth is
increasing.  At higher column densities, the ratio curves turn over (and begin to decrease for some
of the highest \thetaobs bins), as a greater fraction of scattered line photons are absorbed.  The
maximum of the ratio for angle bin~1 (face-on) is $\sim0.3$ and for angle bin~10 (edge-on) is
$\sim0.4$.  This means that the zeroth-order line may contain as little as 71\% of the total line
flux.  

Although the energy interval of the first Compton scattering (6.24--6.4~keV) includes photons from
multiple scatterings, in real data it would be impossible to isolate the once-scattered photons. 
Nevertheless, it is the 6.24--6.4~keV interval that is relevant for the observationally-measured Compton
shoulder, which therefore includes contributions from more than one scattering. Line flux below 6.24 keV
is difficult to distinguish from the continuum. At the lowest column densities, essentially all of the
scattered line photons are in the first-order. We found from our Monte-Carlo results that as \nh
increases, the ratio of the total number of scattered line photons in the 6.24--6.4~keV interval (i.e., the
measured Compton shoulder) to the total number of scattered line photons is $\sim 0.85$, regardless of
\thetaobs and $\Gamma$.  The general behavior of the ratio of the scattered to the zeroth-order line
photon is independent of $\Gamma$.  This is expected as we are considering {\it ratios of numbers of line
photons}, which do not depend on the intrinsic continuum.  Detailed results concerning the shape of the
Compton shoulder will be presented in future work.    

\begin{figure}
\centerline{
\psfig{figure=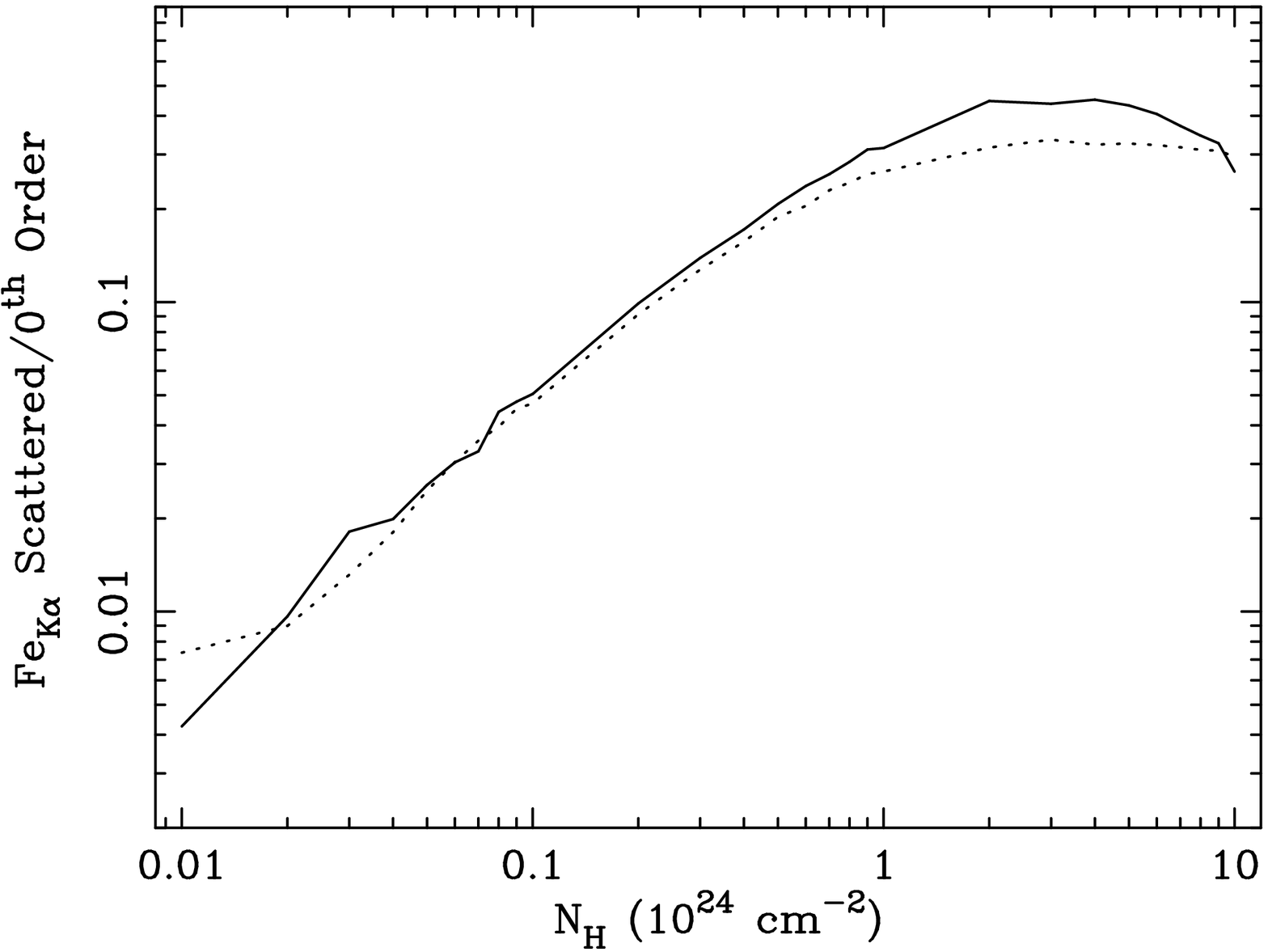,height=8cm}}
\caption{\footnotesize The ratios of the total number of scattered \feka line photons to 
the number of zeroth-order \feka
line photons versus \nhp.  Ratios are shown for an input power-law continuum with $\Gamma=1.9$, for 
 the face-on (bin~1 -- {\it dotted curve}; see \tablecosrangep) and edge-on (bin~10 -- {\it solid curve}; 
 see \tablecosrangep) \thetaobs bins.  Note that for \nh$<10^{23} \rm \ cm^{-2}$, these ratios are 
 affected by small number statistics and should be interpreted with caution.}
\end{figure}

\section{Summary}
\label{torusconcl}

We have calculated Green's functions that may be used to produce spectral-fitting routines to model the
putative neutral toroidal X-ray reprocessor in AGNs for an arbitrary input spectrum.  The absolute size of
the structure does not affect the reprocessed spectrum so our results can be used to model {\it any}
distant-matter, cold, toroidal structure in the central engine, from the optical broad-line region (BLR) and
beyond.  A doughnut-shaped torus geometry has been employed and the model has been calculated for a single
half-opening angle of $60^{\circ}$ and is valid for column densities in the range $10^{22} \rm \ cm^{-2} \
and \ 10^{25} \rm \ cm^{-2}$.  The upper end of the column density range likely satisfies the majority of
current observational needs but can easily be extended in the future.   Our calculations are  fully
relativistic and valid up to 500~keV if the intrinsic incident spectrum has a cut-off below this energy. For
intrinsic spectra that extend beyond 500~keV we have assessed in detail the range of validity of the
calculations. However, the Green's function approach enables this energy range to be easily extended with
further Monte Carlo runs.  We find that our adopted geometry produces significant flux in the soft X-ray band
even for Compton-thick column densities along edge-on lines of sight. Such soft flux can be observationally
important and could manifest itself as complex absorption in the soft X-ray band.  The Monte-Carlo code
treats the reprocessed continuum as well as the three most important fluorescent emission lines (\fekap,
\fekbp, and \nikap) with sufficient detail such that the results are applicable to observations obtained by
current and planned X-ray instrumentation.   

The reprocessed continuum and fluorescent line emission due to Fe~K$\alpha$, Fe~K$\beta$, and Ni~K$\alpha$
are treated self-consistently in our calculations, eliminating the need for {\it ad~hoc} modeling that is
currently common practice. We have found that the reflection spectrum from a Compton-thick face-on torus
that subtends a solid angle of $2\pi$ at the X-ray source is factor of $\sim 6$ weaker than that expected
from a Compton-thick, face-on disk, even though the latter subtends the same solid angle. This is a very
significant difference and emphasizes the universal misinterpretation of the so-called ``reflection
fraction'' (defined in terms of the disk geometry) as a literal solid angle (as a fraction of $2\pi$). The
relative strength of the reflection continuum very critically depends on geometry because it is affected
by the angle of reflection integrated over the surface of the reprocessor. Even different toroidal
geometries give different amplitudes of the reflected spectrum for the same solid angle subtended at the
X-ray source. A direct comparison of our Compton-thick reflected continua with the disk-reflection model
{\tt pexrav}, shows that not only is the amplitude substantially different, but the shape of the soft X-ray
spectrum and the shape and relative amplitude of the $\sim 10-30$~keV ``Compton-hump'' are different
enough to have observational consequences. Since the Fe~K$\alpha$ line EW is physically related to the
reflection continuum, it is also correspondingly weaker than the  Fe~K$\alpha$ line EW expected from a
disk. For our Compton-thick torus observed face-on, the Fe~K$\alpha$ line EW can be as little as $\sim
30$~eV.   Curves of EW of the \feka line versus \nh such as those presented in this work can be generated
for arbitrary input spectra and can be used as an important diagnostic of the reprocessor.  The effects of
geometry and inclination become apparent at column densities larger than $\sim2\times10^{23} \rm \
cm^{-2}$.  Our calculations are detailed enough to allow orientation effects to be modeled in the X-ray
spectra, providing more quantitative tests of the unified model.

Spectral-fitting routines are generally characterized by a trade-off between speed and model detail. 
A variety of spectral-fitting routines may be constructed based on our results, depending on the
desired speed and application.  By design, these routines could be combined with other models in
spectral fitting packages.    

Acknowledgments \\
The authors thank Shinya Ikeda, Hisamitsu Awaki, and
Yuichi Terashima for fruitful discussions.  We thank Andrzej Zdziarski 
for carefully reviewing the manuscript and for constructive comments.
Partial support from NASA grants NNG04GB78A (KM, TY) and
NNX09AD01G (TY) is acknowledged.

\bsp
\label{lastpage}

\end{document}